\documentclass[twocolumn, secnumarabic, amssymb, nobibnotes, prl, superscriptaddress, longbibliography]{revtex4}

\usepackage{graphicx}
\usepackage{amsmath}
\usepackage{float}
\usepackage{color}
\usepackage[english]{babel}
\usepackage{array}
\usepackage[T1]{fontenc}
\usepackage[usenames,dvipsnames]{xcolor}
\usepackage{setspace}
\usepackage{bm}
\usepackage{array} 
\usepackage{makecell} 
\usepackage{lipsum}

\definecolor{darkblue}{rgb}{0,0,0.6}
\definecolor{darkred}{rgb}{0.6,0,0}
\usepackage[colorlinks=true,urlcolor=darkblue,citecolor=darkblue, linkcolor=darkred,hyperfootnotes=false]{hyperref}

\usepackage{tikz}
\usetikzlibrary{arrows.meta}
\usetikzlibrary{decorations.pathreplacing}
\usetikzlibrary{decorations.pathmorphing}
\usetikzlibrary{decorations.markings}

\usetikzlibrary{math}
\usetikzlibrary{tikzmark}
\usepackage{stmaryrd}
\usepackage{stackengine}

\definecolor{color1}{rgb}{0.122,0.467,0.706}
\definecolor{ColorSP}{rgb}{0.925, 0.110, 0.141}
\definecolor{ColorSM}{rgb}{0.259, 0.396, 0.686}
\definecolor{colorFerro}{rgb}{0.957,0.843,0.890}
\definecolor{colorAntiFerro}{rgb}{1.0,0.965,0.835}
\definecolor{colorMixed}{rgb}{1.0,0.902,0.835}
\definecolor{colorPreisach}{rgb}{0.949,0.949,0.949}

\definecolor{color1}{rgb}{0.122,0.467,0.706}
\definecolor{color2}{rgb}{0.839,0.153,0.157}

\definecolor{colorFrozen}{rgb}{0.0,0.75,1.0}
\definecolor{colorSWAP}{rgb}{0.58,0.81,0.58}
\definecolor{newGreenColor}{rgb}{0.24,0.70,0.44}

\definecolor{colorCO_new}{rgb}{0.337243401759531,1.0,0.6304985337243401}
\definecolor{colorWW_new}{rgb}{0.9132231404958671,0.0033670033670029076,0.0}

\definecolor{colorOpen}{rgb}{0.588,0.201,0.541}
\definecolor{colorClosed}{rgb}{0.0,0.039,0.572}

\colorlet{myred}{red!70!black}
\colorlet{xcol}{blue!70!black}

\tikzset{>=latex} 
\tikzstyle{mass}=[line width=0.6,red!30!black,fill=red!40!black!10,rounded corners=1,
                  top color=red!40!black!20,bottom color=red!40!black!10,shading angle=20]
\tikzstyle{force}=[->,myred,very thick,line cap=round]
\tikzstyle{CM}=[red!40!black,fill=red!80!black!80]
\tikzstyle{rope}=[brown!70!black,very thick,line cap=round]
\tikzstyle{myarr}=[-{Latex[length=3,width=2]},thin]

\definecolor{darkblue}{rgb}{0,0,0.6}
\definecolor{darkred}{rgb}{0.6,0,0}

\makeatletter
\renewcommand*{\fnum@figure}{{\normalfont \small{FIG.}~\thefigure}}
\makeatother

\makeatletter
\def\@seccntformat#1{\csname the#1\endcsname\quad}

\renewcommand\paragraph{\theparagraph.\arabic{paragraph}}
\makeatother

\begin{document}

\title{Reentrant transition to collective actuation in active solids with a polarizing field}

\author{Paul Baconnier}
\affiliation{AMOLF, 1098 XG Amsterdam, The Netherlands.}
\affiliation{Gulliver UMR CNRS 7083, ESPCI Paris, PSL Research University, 75005 Paris, France.}
\author{Mathéo Aksil}
\affiliation{Gulliver UMR CNRS 7083, ESPCI Paris, PSL Research University, 75005 Paris, France.}
\author{Vincent Démery}
\affiliation{Gulliver UMR CNRS 7083, ESPCI Paris, PSL Research University, 75005 Paris, France.}
\affiliation{Univ Lyon, ENSL, CNRS, Laboratoire de Physique, F-69342 Lyon, France.}
\author{Olivier Dauchot}
\affiliation{Gulliver UMR CNRS 7083, ESPCI Paris, PSL Research University, 75005 Paris, France.}

\begin{abstract}
Collective actuation in active solids - the spontaneous coherent excitation of a few vibrational modes - emerges from a feedback between structural deformations and the orientation of active forces. It is an excellent candidate as a basic mechanism for oscillatory dynamics and regulation in dense living systems, and a better control over its onset would open new avenues in the life sciences.
Combining model experiments, simulations and theory, we study the dynamics of such an active solid in the presence of an external field that polarizes the active forces.
The experiments reveal a novel oscillatory regime absent at zero field. The theoretical analysis of a single agent demonstrates that the small field oscillations and the large field ones can be  mapped onto the bounded and unbounded phase dynamics of a nonlinear pendulum.
In the many agents case, the transition to collective actuation is promoted at low field, leading to a reentrant transition. 
\end{abstract}

\maketitle

Active solids -- dense assemblies or elastic structures made of active units -- encompass a wide class of systems ranging from biological \cite{koenderink2009active, serra2012mechanical, deforet2014emergence, prost2015active, bull2021ciliary, tan2022odd} to man-made materials \cite{ferrante2013elasticity, briand2016crystallization, briand2018spontaneously, li2019particle, zheng2023self, xi2024emergent, veenstra2024non, veenstra2025adaptive} to theoretical models \cite{henkes2011active, menzel2013traveling, berthier2013non, bi2016motility, woodhouse2018autonomous, giavazzi2018flocking, janssen2019active, maitra2019oriented, klongvessa2019active, ronceray2019stress, scheibner2020odd, canavello2024polar}.
They are particularly relevant when studying dense biological systems, like confined cell monolayers \cite{petrolli2019confinement, peyret2019sustained}, dense bacterial suspensions \cite{chen2017weak, liu2021viscoelastic}, and dense pedestrian crowds \cite{Gu2025}.
In sharp contrast with active liquids, their positional degrees of freedom have a reference configuration.

Collective actuation, the spontaneous coherent excitation of a few vibrational modes, was first reported in the numerical study of a dense packing of soft active particles~\cite{henkes2011active}, then reported in two very different experimental systems, one composed of freely rotating self-propelled particles located at the nodes of a spring network pinned at its boundaries~\cite{baconnier2022selective}, the other being a bacterial biofilm \cite{xu2022autonomous}. In both cases the key role of a non-linear elasto-active feedback between the deformations of the structure and the orientations of the active units was demonstrated~\cite{baconnier2022selective, xu2022autonomous, baconnier2023discontinuous, baconnier2024noise}. This feedback takes its origin from the reorientation of individual active units on their velocities, a generic process referred to as self-alignment \cite{baconnier2025self}.

Living active systems also have the ability to respond to various types of environmental cues and can polarize towards or away from these signals, e.g., by chemotaxis or galvanotaxis \cite{sengupta2021principles}. This polarization drives important biological processes such as wound healing \cite{sengupta2021principles, kennard2020osmolarity}, immune responses \cite{sun2019infection}, and morphogenesis \cite{lecuit2007cell, farge2011mechanotransduction, miller2013interplay, goodwin2021mechanics}. 

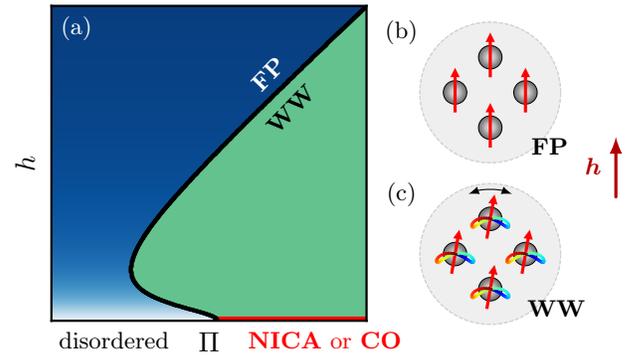
\begin{figure}[t!]
\vspace*{-0.6cm}
\hspace*{-1.1cm}
\begin{tikzpicture}

\node[rotate=0] at (5.7,-5.9) {\includegraphics[height=5.4cm]{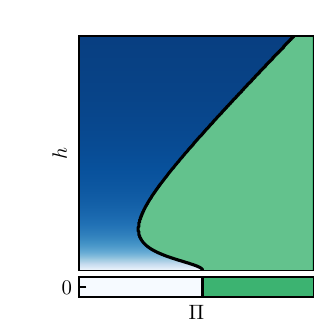}};

\node[rotate=48] at (7.19,-4.82) {\small \textbf{WW}};
\node[rotate=48] at (6.89,-4.4) {\small \color{white} \textbf{FP}};

\node at (5.33,-7.90) {\footnotesize disordered};
\node at (7.23,-7.90) {\footnotesize \color{black} \textbf{NICA} \  \textbf{CO}};
\node at (7.43,-7.90) {\tiny \color{black} /};

\node at (3.85,-4.06) {\small (a)};

\node[rotate=0] at (10.0,-4.95) {\includegraphics[height=2.6cm]{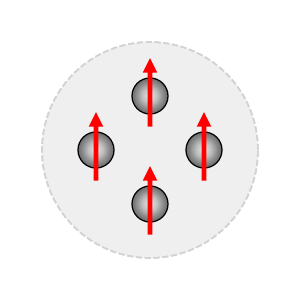}};
\node[rotate=0] at (10.0,-7.1) {\includegraphics[height=2.6cm]{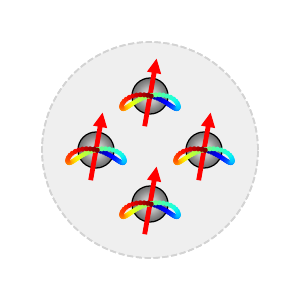}};

\draw[<->] (10.0,-7.1) ++(70:0.85) arc[start angle=70, end angle=110, radius=0.85];

\draw[force] (11.67,-6.34) -- (11.67,-5.54);
 \node[rotate=0] at (11.37,-5.93) {\small \color{myred} $\boldsymbol{h}$};

\node at (8.82,-4.15) {\small (b)};
\node at (8.82,-6.3) {\small (c)};

\node at (10.79,-5.67) {\small \textbf{FP}};
\node at (10.88,-7.82) {\small \textbf{WW}};

\end{tikzpicture}
\vspace*{-0.57cm}
\caption{\small{Active solids in a polarizing field. (a) Schematic phase diagram as a function of the elasto-active coupling $\Pi$ and the field amplitude $h$. At $h = 0$, a transition between a disordered regime and collective actuation – either noise induced (NICA), or chiral oscillations (CO) – takes place \cite{baconnier2022selective, baconnier2024noise}. Adding a field, the disordered phase polarizes (FP) (color-coded from light to dark blue as polarization increases). In the green region, oscillating dynamics take place, and for large enough fields, a new dynamical regime emerges, taking the form of bounded oscillations around the field orientation, analogous to the motion of Windscreen Wipers (WW). (b-c) Schematic FP and WW regimes; red arrows: polarities $\boldsymbol{\hat{n}}_i$; trajectories of particle positions color-coded from blue to red.}}
\label{fig:fig1}
\end{figure}

In this Letter, we investigate the largely unexplored effect of an external polarizing field on the collective dynamics of active solids combining model experiments, simulations of an agent-based model, and theory.  More specifically, we consider networks of freely rotating self-propelled particles connected by elastic springs with various boundary conditions. When the experimental plane is tilted, the active units orient opposite to the gravity force. We take advantage of this mechanism to impose a homogeneous polarizing field to the system, and obtain the generic phase diagram reported on Fig. \ref{fig:fig1}a.
In the zero-field case, reported previously~\cite{baconnier2022selective,baconnier2024noise}, a transition takes place for large enough elasto-active coupling $\Pi$, to be defined below, from a disordered regime, where the orientation of each agent diffuses randomly, to a collective actuation regime, where the dynamics condensates on two vibrational modes and performs regular oscillations.
The selection of the excited modes relies on their softness and shape, and is therefore governed by the lattice structure and boundary conditions.
When two degenerate modes are selected—\cite{baconnier2022selective}, these oscillations take the form of Chiral Oscillations (CO). In contrast, when the lowest energy mode is strongly gapped from the higher energy ones, the collective actuation takes the form of a back-and-forth oscillation along the softest mode and requires the presence of noise, hence the name Noise-Induced Collective Actuation (NICA)~\cite{baconnier2024noise}. Here we show that adding a field, (i) the disordered phase turns into a Frozen Polarized phase (FP), (ii) a new dynamical regime emerges where the active units perform bounded and synchronized oscillations around the field orientation, analogous to the motion of Windscreen Wipers (WW), (iii) as the field amplitude increases, a reentrant transition occurs between the FP and WW regimes, to be understood as a purely collective effect.

The experimental set-up consists of elastic structures, composed of $N$ active units connected by springs of stiffness $k$ and rest length $l_0$~ \cite{baconnier2022selective, baconnier2024noise}.
Each unit is made of a Hexbug$\copyright$, a centimetric battery-powered running robot, embedded in a $3$D printed annulus.
The active unit $i$ exerts a force $F_0 \boldsymbol{\hat{n}}_i$ along its orientation $\boldsymbol{\hat{n}}_i$,
and is displaced from its reference position by the vector $\boldsymbol{u}_i$. Each unit is free to rotate: $\boldsymbol{\hat{n}}_i$ diffuses angularly due to the mechanical noise inherent to the agent design, and reorients towards its velocity $\dot{\boldsymbol{u}}_i$ over a length $l_a$, the alignment length, an effect which has been called self-alignment \cite{baconnier2022selective, baconnier2025self}.
The strength of this coupling between displacements and polarities is set by the ratio $\Pi = l_e/l_a$, with $l_e = F_0/k$; it is varied by using springs of two different stiffnesses (\textit{soft} and \textit{stiff}) and adjusting their length (Supplemental Material). Moreover, we impose a spring extension through the boundary condition $\alpha = l_\mathrm{eq}/l_0$, where $l_\mathrm{eq}$ is the equilibrium length of the springs in the reference configuration under tension.
The specificity of this work is to tilt the plane of the experiment by an angle $\beta \in \left[ 0^{\circ}, 21.4^{\circ} \right]$. Because the mass of the Hexbug is not distributed evenly along the body’s axis, a torque reorients the active unit, acting as a polarizing field of amplitude $h \propto g \sin \beta$, in the direction opposite to the gravity force, which we denote $\boldsymbol{\hat{e}}_{\parallel}$.
In the following, we denote $\theta_i$ (resp. $\varphi_i$) the angle between $\boldsymbol{\hat{e}}_{\parallel}$ and $\boldsymbol{\hat{n}}_i$ (resp. $\boldsymbol{u}_i$).

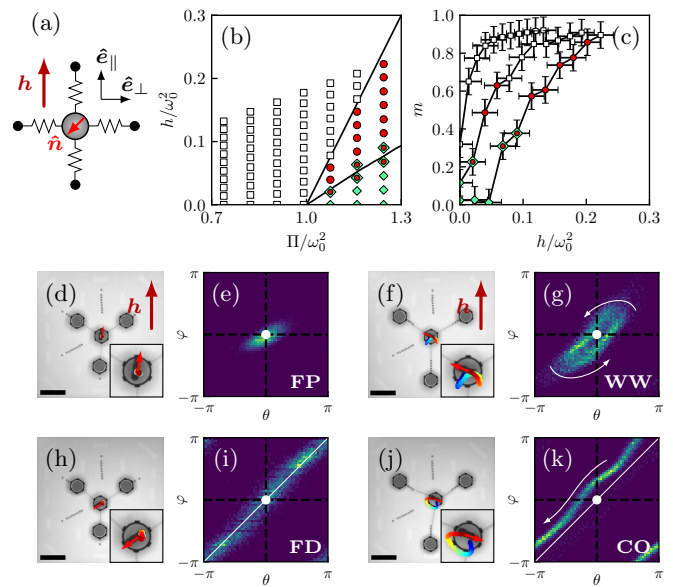
\begin{figure}[t!]
\vspace*{-0.3cm}
\hspace*{-0.25cm}
\begin{tikzpicture}

\node[rotate=0] at (-0.2,-0.7) {\includegraphics[height=1.32cm]{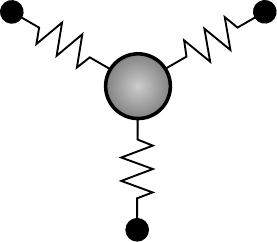}};

\draw[->, thick, color=red] (-0.09,-0.39) -- (-0.33,-0.63);
\node[rotate=0] at (-0.48,-0.73) {\footnotesize \color{red} $\boldsymbol{\hat{n}}$};

\draw[force] (-0.63,0.16) -- (-0.63,0.81);
\node[rotate=0] at (-0.88,0.47) {\footnotesize \color{myred} $\boldsymbol{h}$};

\draw[->] (0.13,0.25) -- (0.13,0.65);
\draw[->] (0.13,0.25) -- (0.53,0.25);

\node[rotate=0] at (0.22,0.78) {\footnotesize $\boldsymbol{\hat{e}}_{\parallel}$};
\node[rotate=0] at (0.59,0.44) {\footnotesize $\boldsymbol{\hat{e}}_{\perp}$};

\node[rotate=0] at (-0.6,1.25) {\small (a)};

\node[rotate=0] at (2.5,0.0) {\includegraphics[height=3.6cm]{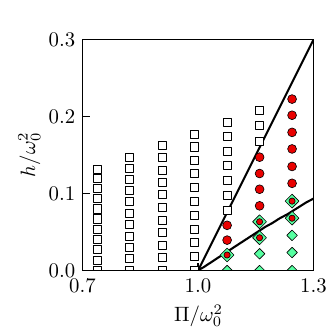}};

\node[rotate=0] at (5.8,0.0) {\includegraphics[height=3.6cm]{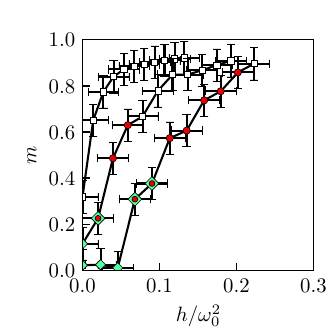}};

\node[rotate=0] at (1.93,1.07) {\small (b)};
\node[rotate=0] at (7.11,1.07) {\small (c)};

\node[rotate=0] at (0.12,-2.88) {\includegraphics[height=1.67cm]{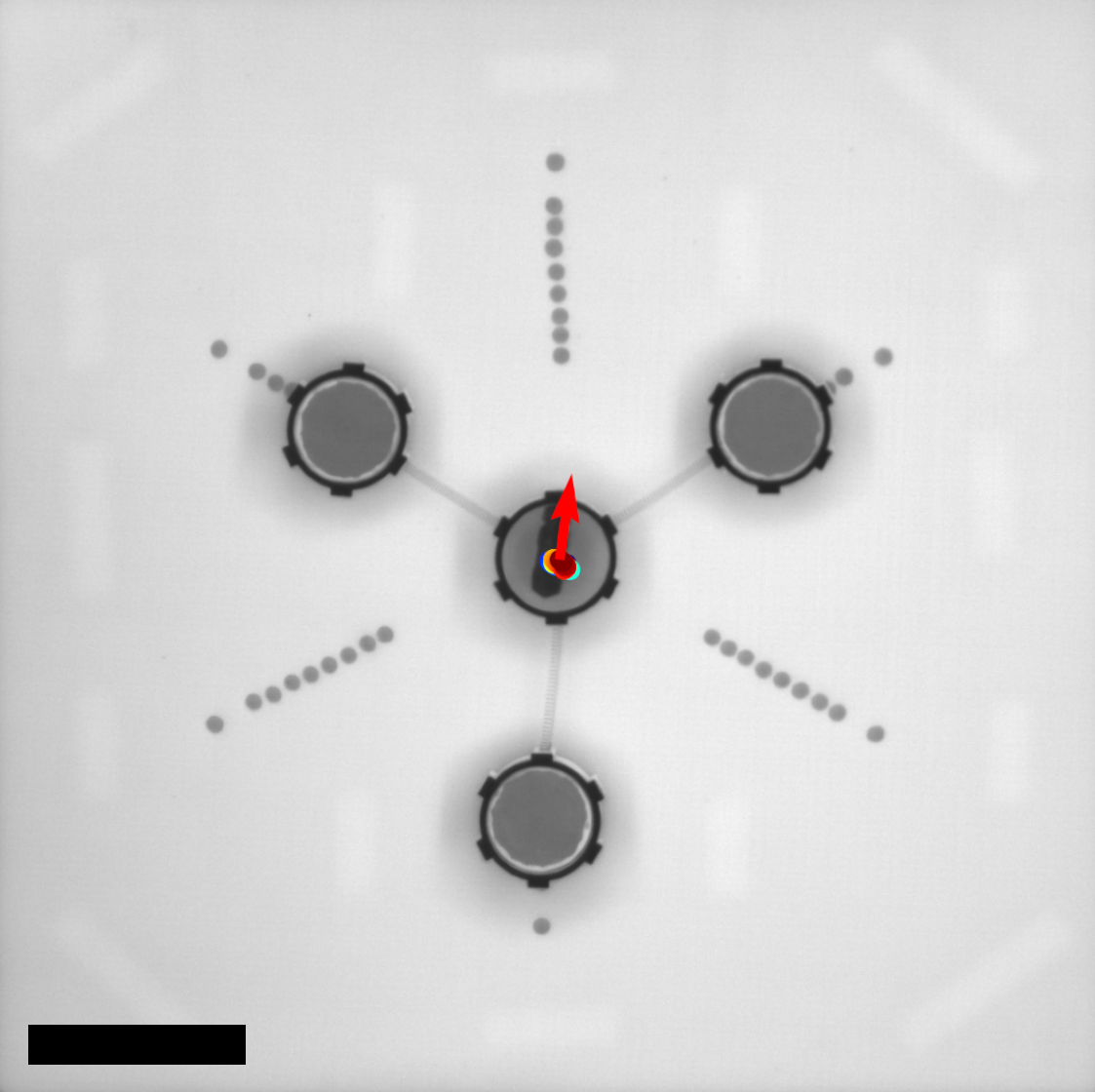}};
\node[rotate=0, anchor=south east] at (1.08,-3.84) {\includegraphics[height=0.7cm]{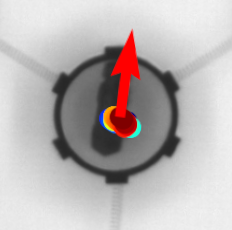}};
\draw[] (0.95,-3.71) rectangle (0.25,-3.01);

\node[rotate=0] at (2.2,-3.0) {\includegraphics[height=2.2cm]{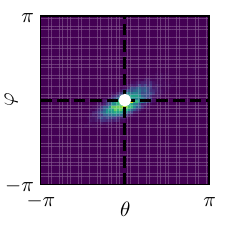}};

\node[rotate=0] at (4.52,-2.88) {\includegraphics[height=1.67cm]{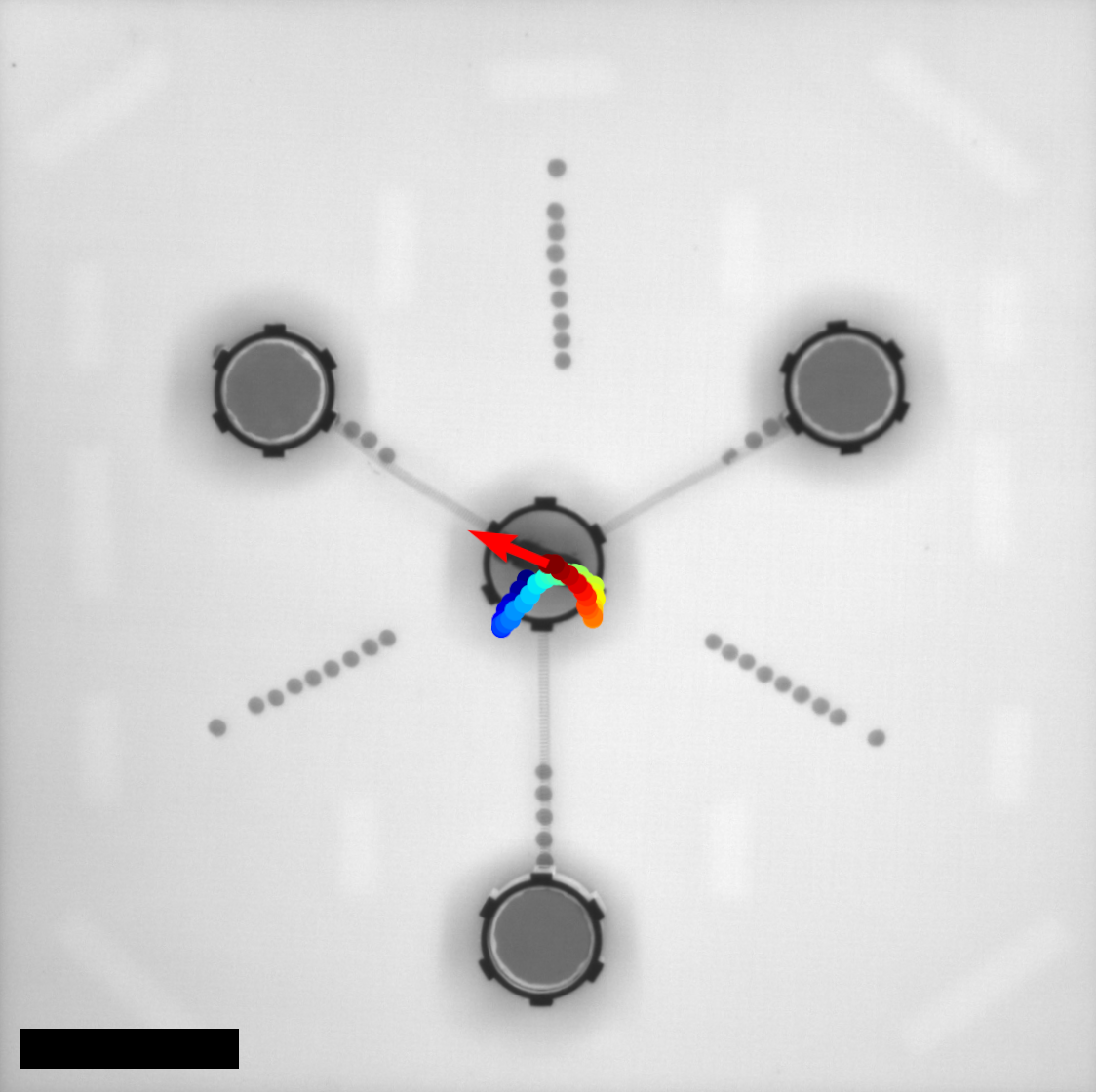}};
\node[rotate=0, anchor=south east] at (5.48,-3.84) {\includegraphics[height=0.7cm]{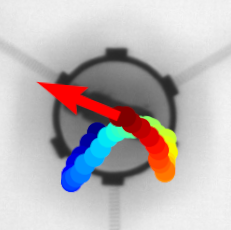}};
\draw[] (5.35,-3.71) rectangle (4.65,-3.01);

\node[rotate=0] at (6.6,-3.0) {\includegraphics[height=2.2cm]{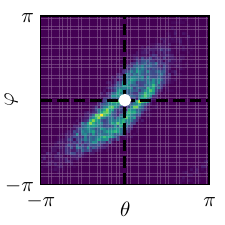}};

\node[rotate=0] at (0.12,-5.08) {\includegraphics[height=1.67cm]{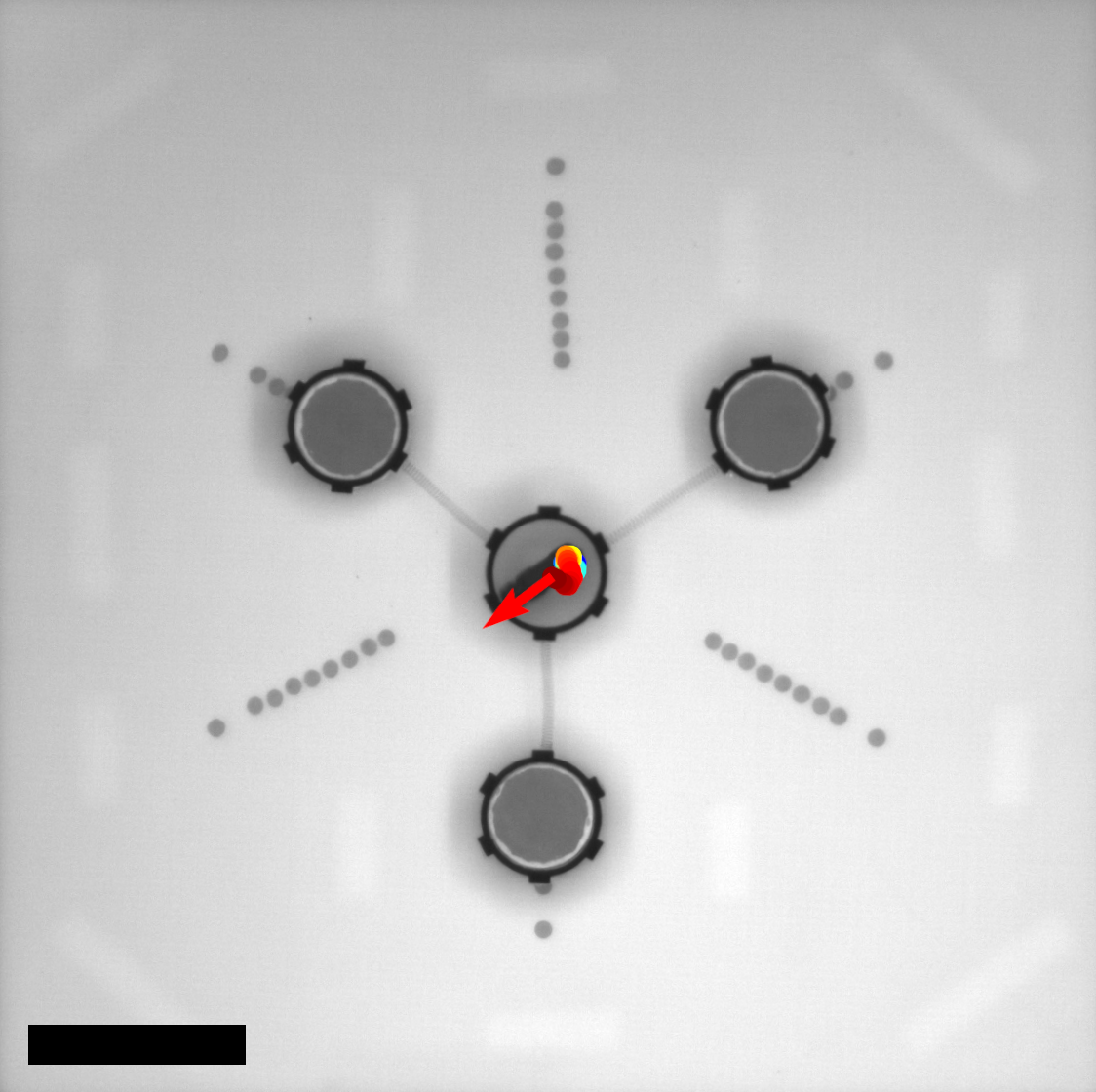}};
\node[rotate=0, anchor=south east] at (1.08,-6.04) {\includegraphics[height=0.7cm]{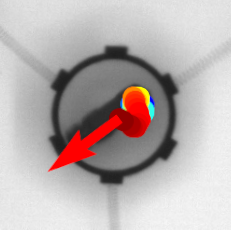}};
\draw[] (0.95,-5.91) rectangle (0.25,-5.21);

\node[rotate=0] at (2.2,-5.2) {\includegraphics[height=2.2cm]{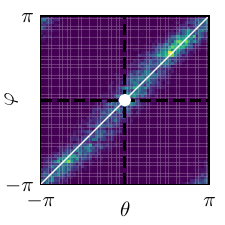}};

\node[rotate=0] at (4.52,-5.08) {\includegraphics[height=1.67cm]{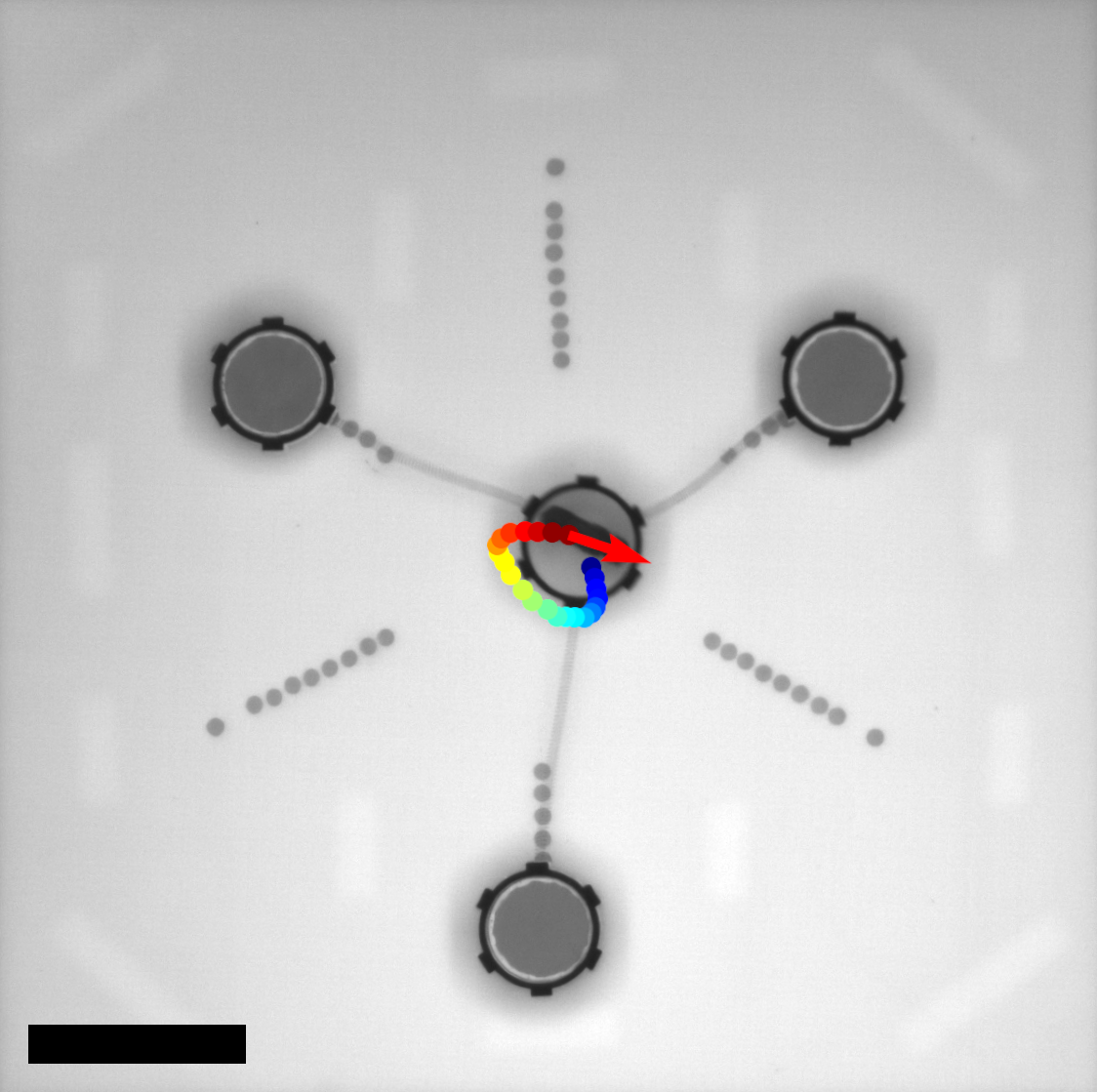}};
\node[rotate=0, anchor=south east] at (5.48,-6.04) {\includegraphics[height=0.7cm]{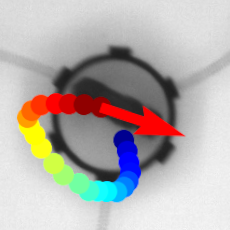}};
\draw[] (5.35,-5.91) rectangle (4.65,-5.21);

\node[rotate=0] at (6.6,-5.2) {\includegraphics[height=2.2cm]{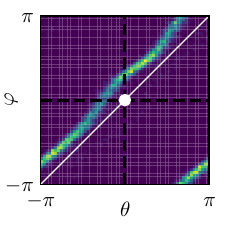}};

\node[rotate=0] at (2.85,-3.5) {\scriptsize \color{white} \textbf{FP}};
\node[rotate=0] at (7.14,-3.5) {\scriptsize \color{white} \textbf{WW}};

\node[rotate=0] at (2.85,-5.7) {\scriptsize \color{white} \textbf{FD}};
\node[rotate=0] at (7.21,-5.7) {\scriptsize \color{white} \textbf{CO}};

\node[rotate=0] at (-0.4,-2.35) {\small (d)};
\node[rotate=0] at (3.98,-2.35) {\small (f)};

\node[rotate=0] at (1.79,-2.35) {\small \color{white} (e)};
\node[rotate=0] at (6.2,-2.35) {\small \color{white} (g)};

\node[rotate=0] at (-0.4,-4.55) {\small (h)};
\node[rotate=0] at (3.98,-4.55) {\small (j)};

\node[rotate=0] at (1.79,-4.55) {\small \color{white} (i)};
\node[rotate=0] at (6.2,-4.55) {\small \color{white} (k)};

\draw[force] (0.8,-2.80) -- (0.8,-2.15);
\node[rotate=0] at (0.57,-2.49) {\footnotesize \color{myred} $\boldsymbol{h}$};

\draw[force] (5.18,-2.80) -- (5.18,-2.15);
\node[rotate=0] at (4.95,-2.49) {\footnotesize \color{myred} $\boldsymbol{h}$};

\draw[-{Latex[length=3,width=2]},thin,white] (6.85,-4.55) to[out=210,in=35] (6.0,-5.4);

\draw[-{Latex[length=3,width=2]},thin,white] (7.24,-2.56) to[out=-245,in=45] (6.54,-2.66);
\draw[-{Latex[length=3,width=2]},thin,white] (6.1,-3.3) to[out=295,in=-135] (6.9,-3.2);

\end{tikzpicture}
\vspace*{-0.9cm}
\caption{\small{Single active unit experiments. (a) Geometry and notations; (b) Phase diagram (white square: FP, red circles: WW, green diamonds: CO, red circles inside green diamonds: coexistence between WW and CO, the black lines indicate $\Pi_c = \omega_{0}^{2} + h$ and $\Pi^{\star} = \omega_{0}^{2} + 3h$). (c) Polarization $m$ as a function of $h/\omega_{0}^{2}$ for $\Pi/\omega_{0}^{2} \in [0.74, 1.08, 1.24]$ (same markers as (b)); (d/f/h/j) Real space dynamics of the FP, WW, FD, CO regimes (red arrows: polarity $\boldsymbol{\hat{n}}$, trajectories of particle positions color-coded from blue to red, scale bars: $10$ cm). (e/g/i/k) Probability densities $\rho(\theta, \varphi)$, the white arrows indicate the direction of the dynamics when relevant. The parameter values are: (d/e) FP: $\Pi/\omega_{0}^{2} = 0.91$, $h/\omega_{0}^{2} = 0.08$ ($\beta = 10.7^{\circ}$); (f/g) WW: $\Pi/\omega_{0}^{2} = 1.24$, $h/\omega_{0}^{2} = 0.11$ ($\beta = 10.7^{\circ}$); (h/i) FD: $\Pi/\omega_{0}^{2} = 0.91$, $h/\omega_{0}^{2} = 0.0$ ($\beta = 0^{\circ}$); (j/k) CO: $\Pi/\omega_{0}^{2} = 1.24$, $h/\omega_{0}^{2} = 0.0$ ($\beta = 0^{\circ}$).}}
\label{fig:fig2}
\end{figure}

Some of the physics connecting activity, elasticity, and the external field can already be captured at the level of a single active unit connected to the three static vertices of a regular triangle (Fig.~\ref{fig:fig2}a).
This elastic structure has two degenerate normal modes along $\boldsymbol{\hat{e}}_{\perp}$ and $\boldsymbol{\hat{e}}_{\parallel}$ of stiffness $\omega_0^2$.
The overdamped harmonic dynamics of the agent is described by the dimensionless equations
\begin{subequations} \label{eq:sp}
\begin{align}
\dot{\boldsymbol{u}} &= \Pi \boldsymbol{\hat{n}} - \omega_{0}^{2} \boldsymbol{u}, \label{eq1:sp} \\
\dot{\boldsymbol{\hat{n}}} &= ( \boldsymbol{\hat{n}} \times \left[ \dot{\boldsymbol{u}} + \boldsymbol{h} \right] ) \times \boldsymbol{\hat{n}} + \sqrt{2D} \boldsymbol{\hat{n}_\perp}, \label{eq2:sp}
\end{align}
\end{subequations}
where the first equation describes the motion resulting from the balance of active and elastic forces, and the second equation accounts for the self-alignment, the polarization by the field $\boldsymbol{h}$, and the angular noise. The time unit is the ratio of the damping coefficient to the spring stiffness $k$ and the length unit is the alignment length $l_a$.

In the absence of an external field \cite{dauchot2019dynamics, baconnier2022selective}, a drift-pitchfork bifurcation takes place when $\Pi$ increases. For $\Pi < \omega_0^2$, the active unit diffuses along the circle of radius $|\boldsymbol{u}| = R = \Pi/\omega_0^2$ with $\theta = \varphi$ (Figs. \ref{fig:fig2}h-i): this is the Frozen-Disordered (FD) regime.
When $\Pi > \omega_0^2$, the fixed points of the noiseless dynamics composing this circle become unstable and a finite angle $\gamma = \theta - \varphi$ drives the system along a limit cycle of radius $R = (\Pi/\omega_0^2)^{1/2}$, at a rotation rate $\Omega = \pm \omega_0 (\Pi - \omega_0^2)^{1/2}$ (Figs. \ref{fig:fig2}j-k): this is the CO regime.
Adding a field breaks the rotational symmetry responsible for the degeneracy of the fixed points.
For $\Pi < \omega_0^2$, the system enters the Frozen Polarized (FP) regime (Figs. \ref{fig:fig2}d-e): the distribution of orientations peaks around the direction of the field, and the average polarization $m = \left| \int_0^{2\pi} e^{i\theta} \rho(\theta) d\theta \right|$ increases with $h$ (Fig. \ref{fig:fig2}c).
For $\Pi > \omega_0^2$, three regimes are observed depending on the field amplitude (Movies $1$ to $3$).
At small fields, the CO regime subsists, with a temporal modulation of the angle $\gamma$ at the CO rotation frequency. For intermediate fields, a new dynamical regime emerges, where the orientation of the active unit oscillates around that of the field (Figs. \ref{fig:fig2}f-g), leading to a back-and-forth motion analogous to that of Windscreen Wipers (WW).
Eventually, large fields stabilize the FP regime and delay the transition to CO.
The transitions between the FP and the WW regimes and between the WW and the CO regimes appear close to  $\Pi_c = \omega_0^2 + h$ and $\Pi^{\star} = \omega_0^2 + 3h$, respectively (Fig. \ref{fig:fig2}b).

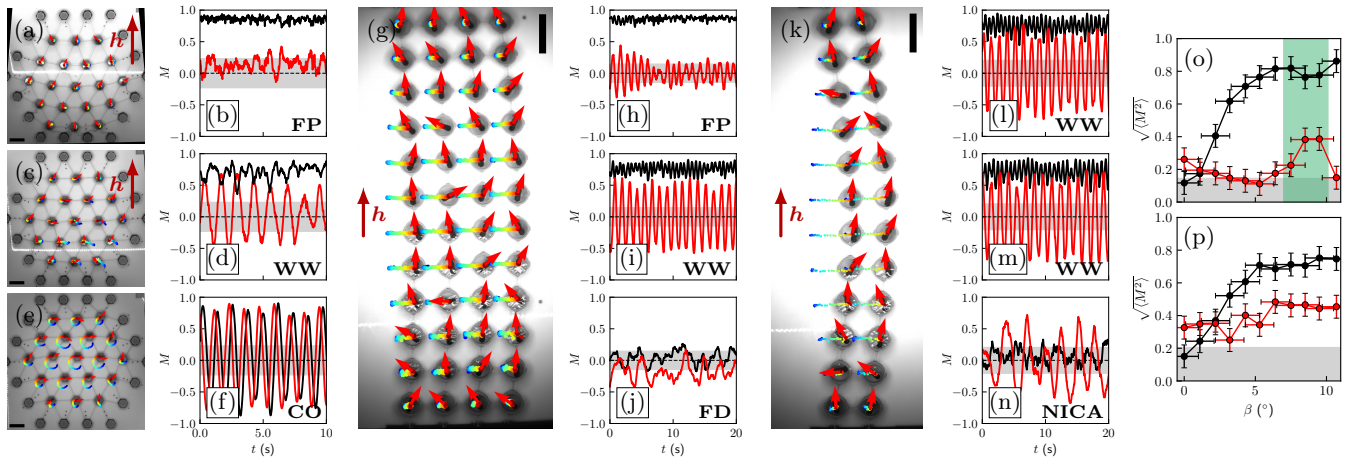
\begin{figure*}[t!]
\hspace*{-0.13cm}
\begin{tikzpicture}

\draw[color=white] (-0.95,-1.2) rectangle (17.0,4.8);

\node[rotate=0] at (0.0,0.0) {\includegraphics[height=1.8cm]{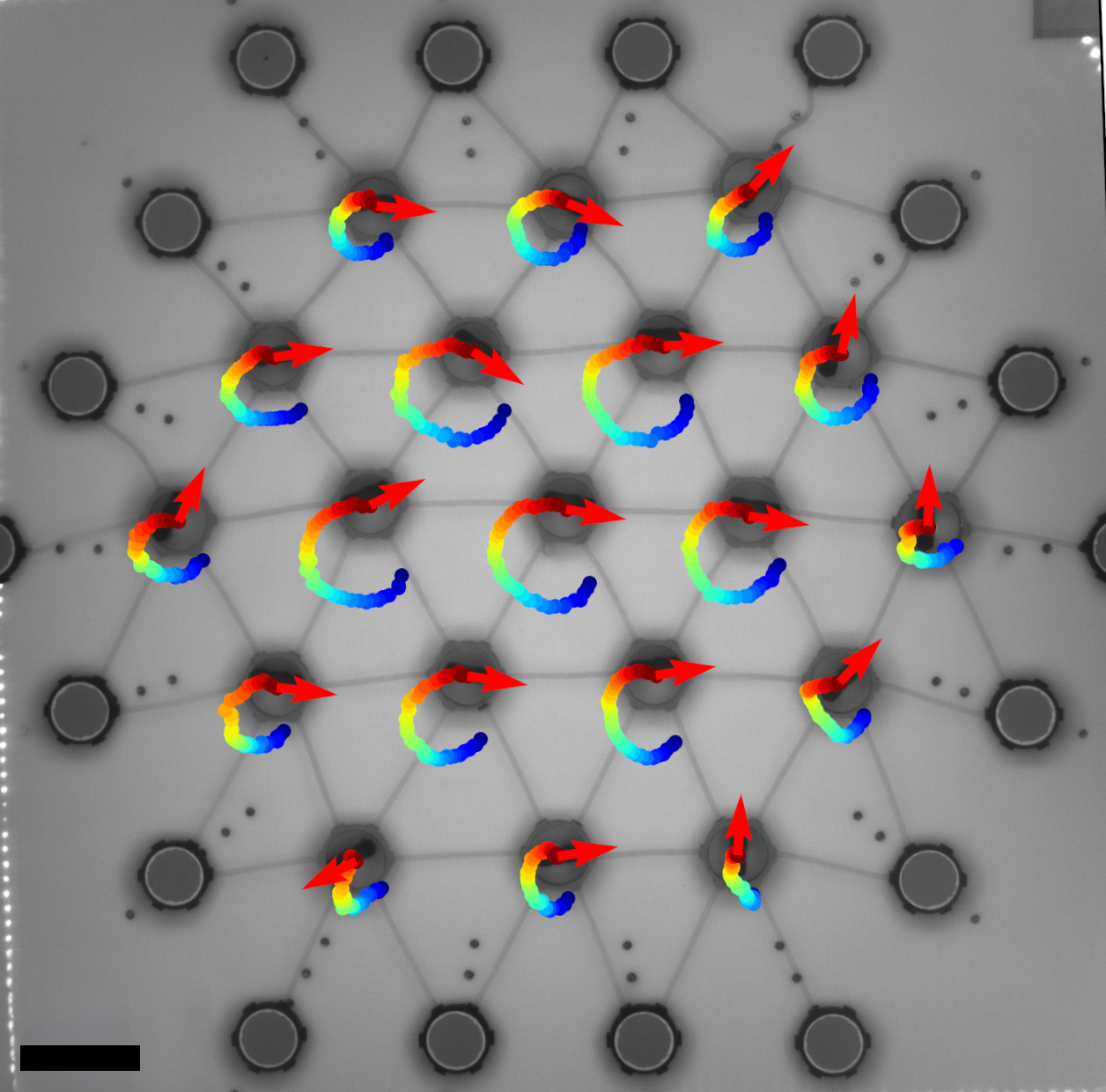}};
\node[rotate=0] at (0.0,1.9) {\includegraphics[height=1.8cm]{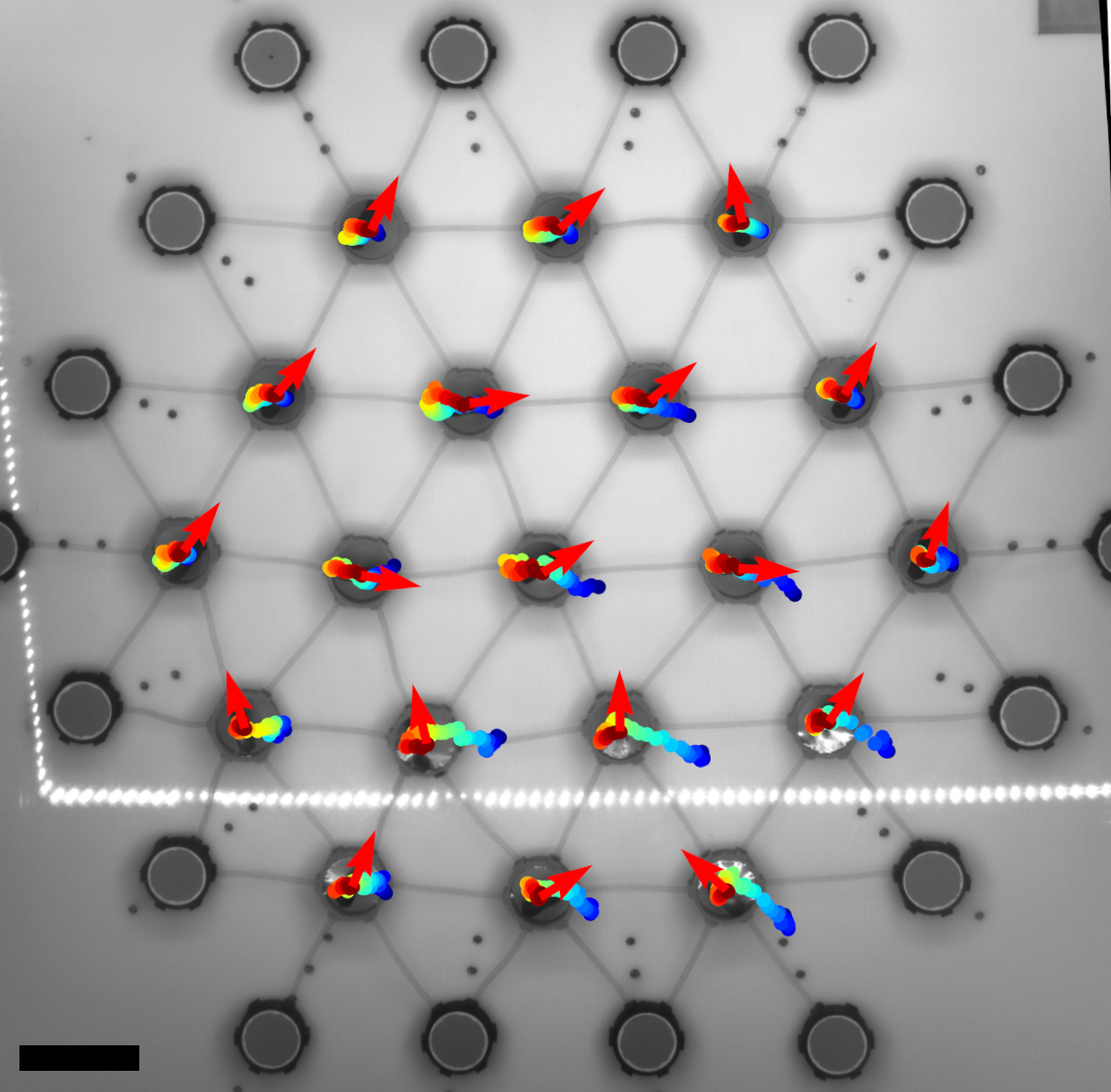}};
\node[rotate=0] at (0.0,3.8) {\includegraphics[height=1.8cm]{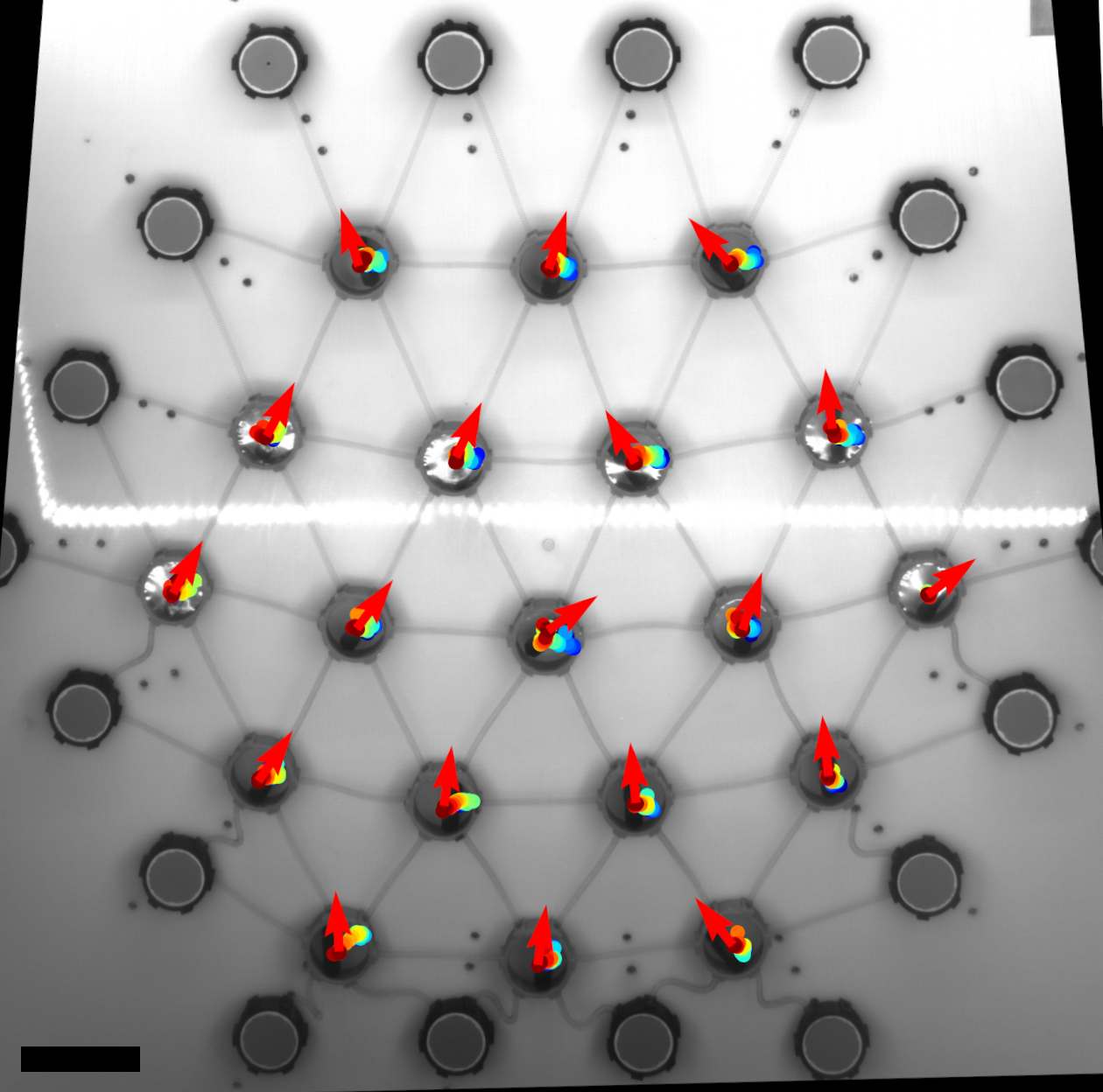}};

\draw[force] (0.75,2.05) -- (0.75,2.70);
\node[rotate=0] at (0.54,2.36) {\footnotesize \color{myred} $\boldsymbol{h}$};

\draw[force] (0.75,3.97) -- (0.75,4.62);
\node[rotate=0] at (0.54,4.28) {\footnotesize \color{myred} $\boldsymbol{h}$};

\fill[white, draw=black, line width=0.05mm] (-0.70-0.21,4.50-0.2) rectangle (-0.70+0.21,4.50+0.2);
\fill[white, draw=black, line width=0.05mm] (-0.70-0.21,2.59-0.2) rectangle (-0.70+0.21,2.59+0.2);
\fill[white, draw=black, line width=0.05mm] (-0.70-0.21,0.695-0.2) rectangle (-0.70+0.21,0.695+0.2);

\node[rotate=0] at (-0.70,4.50) {\small (a)};
\node[rotate=0] at (-0.70,2.59) {\small (c)};
\node[rotate=0] at (-0.70,0.695) {\small (e)};

\node[rotate=0] at (2.2,-0.13) {\includegraphics[height=2.4cm]{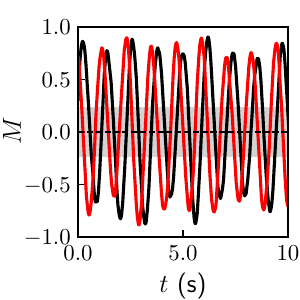}};
\node[rotate=0] at (2.2,1.78) {\includegraphics[height=2.4cm]{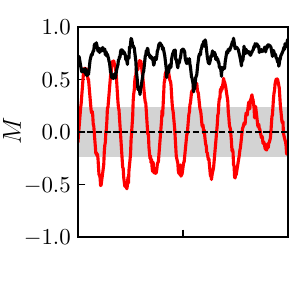}};
\node[rotate=0] at (2.2,3.69) {\includegraphics[height=2.4cm]{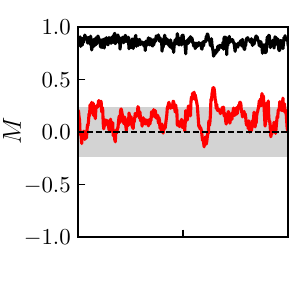}};

\fill[white, draw=black, line width=0.05mm] (1.95-0.21,3.27-0.2) rectangle (1.95+0.21,3.27+0.2);
\fill[white, draw=black, line width=0.05mm] (1.95-0.21,1.36-0.2) rectangle (1.95+0.21,1.36+0.2);
\fill[white, draw=black, line width=0.05mm] (1.95-0.21,-0.55-0.2) rectangle (1.95+0.21,-0.55+0.2);

\node[rotate=0] at (1.95,3.27) {\small (b)};
\node[rotate=0] at (1.95,1.36) {\small (d)};
\node[rotate=0] at (1.95,-0.55) {\small (f)};

\node[rotate=0, anchor=east] at (3.40,3.15) {\scriptsize \textbf{FP}};
\node[rotate=0, anchor=east] at (3.40,1.24) {\scriptsize \textbf{WW}};
\node[rotate=0, anchor=east] at (3.40,-0.67) {\scriptsize \textbf{CO}};

\node[rotate=90] at (5.05,1.91) {\includegraphics[height=2.58cm]{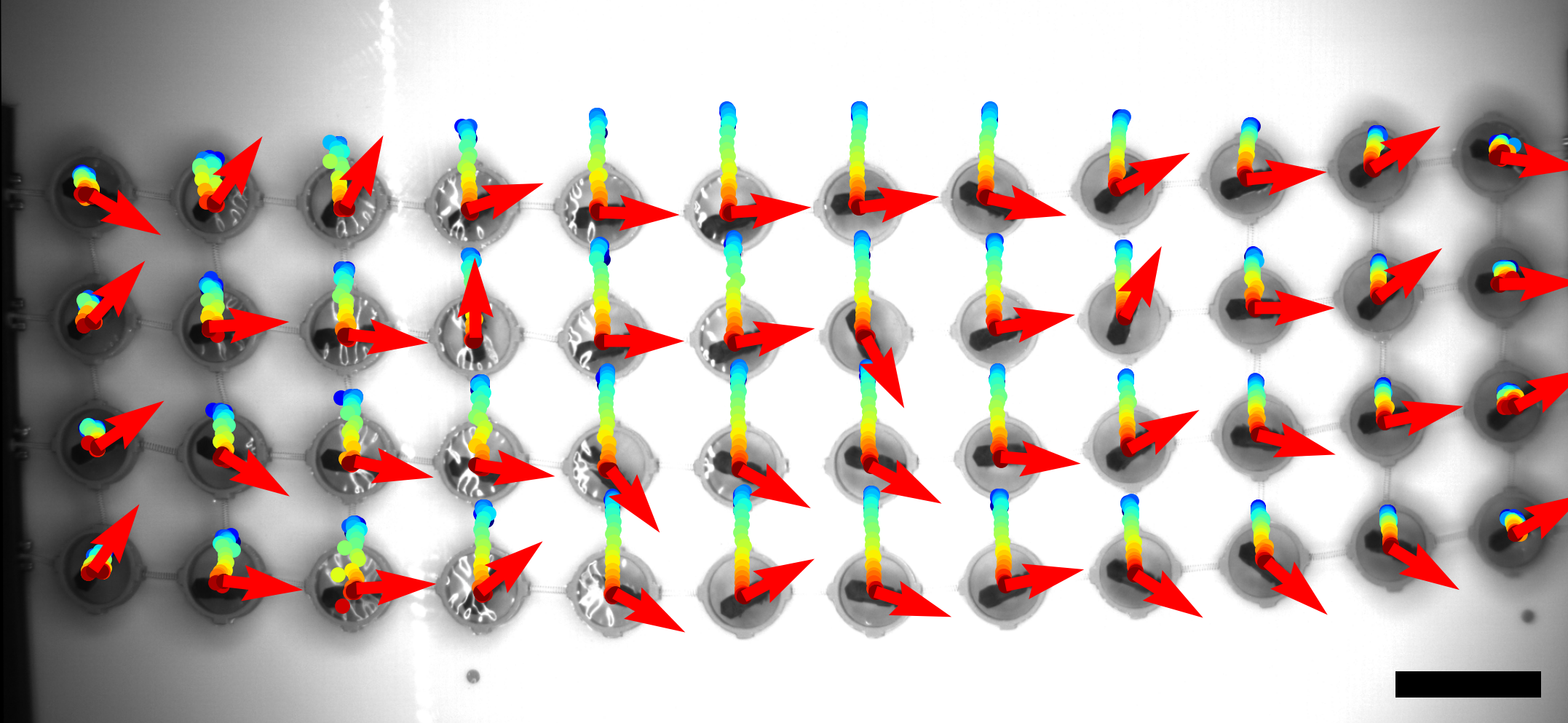}};

\fill[white, draw=black, line width=0.05mm] (3.98-0.21,4.50-0.2) rectangle (3.98+0.21,4.50+0.2);

\node[rotate=0] at (3.98,4.50) {\small (g)};
\draw[force] (3.82,1.67) -- (3.82,2.32);
\node[rotate=0] at (4.03,1.97) {\footnotesize \color{myred} $\boldsymbol{h}$};

\node[rotate=0] at (7.65,-0.13) {\includegraphics[height=2.4cm]{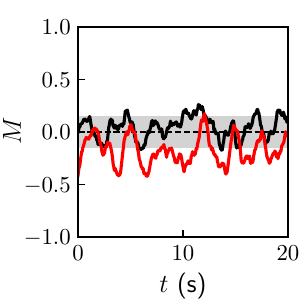}};
\node[rotate=0] at (7.65,1.78) {\includegraphics[height=2.4cm]{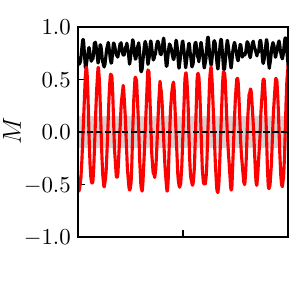}};
\node[rotate=0] at (7.65,3.69) {\includegraphics[height=2.4cm]{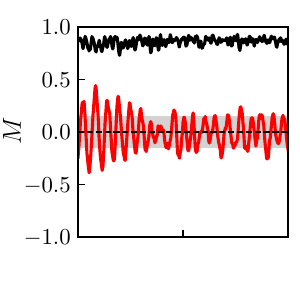}};

\fill[white, draw=black, line width=0.05mm] (7.40-0.21,3.26-0.2) rectangle (7.40+0.21,3.26+0.2);
\fill[white, draw=black, line width=0.05mm] (7.38-0.20,1.35-0.2) rectangle (7.38+0.20,1.35+0.2);
\fill[white, draw=black, line width=0.05mm] (7.39-0.21,-0.56-0.2) rectangle (7.39+0.21,-0.56+0.2);

\node[rotate=0] at (7.40,3.27) {\small (h)};
\node[rotate=0] at (7.38,1.36) {\small (i)};
\node[rotate=0] at (7.39,-0.55) {\small (j)};

\node[rotate=0, anchor=east] at (8.84,3.15) {\scriptsize \textbf{FP}};
\node[rotate=0, anchor=east] at (8.84,1.24) {\scriptsize \textbf{WW}};
\node[rotate=0, anchor=east] at (8.84,-0.67) {\scriptsize \textbf{FD}};

\node[rotate=90] at (10.25,1.91) {\includegraphics[height=2.0cm]{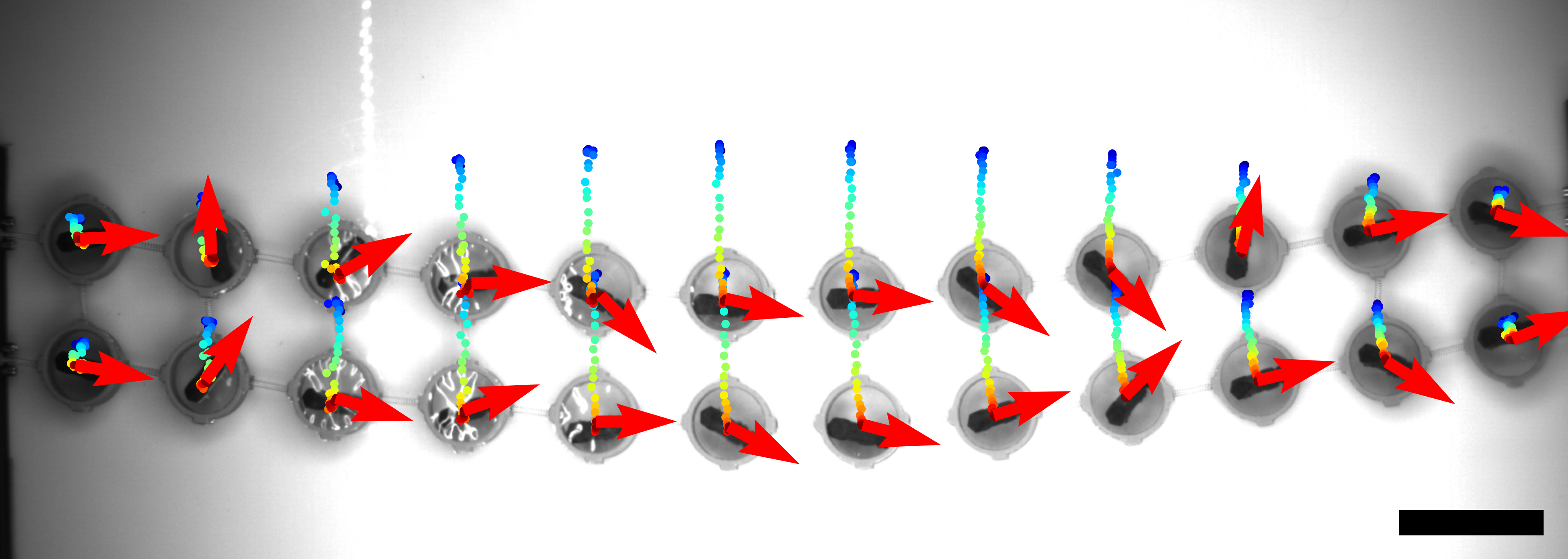}};

\fill[white, draw=black, line width=0.05mm] (9.47-0.21,4.50-0.2) rectangle (9.47+0.21,4.50+0.2);

\node[rotate=0] at (9.47,4.5) {\small (k)};
\draw[force] (9.37,1.67) -- (9.37,2.32);
\node[rotate=0] at (9.58,1.97) {\footnotesize \color{myred} $\boldsymbol{h}$};

\node[rotate=0] at (12.6,-0.13) {\includegraphics[height=2.4cm]{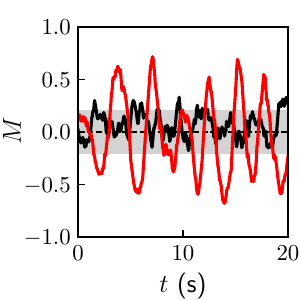}};
\node[rotate=0] at (12.6,1.78) {\includegraphics[height=2.4cm]{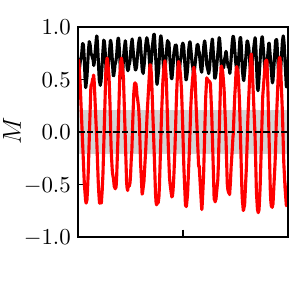}};
\node[rotate=0] at (12.6,3.69) {\includegraphics[height=2.4cm]{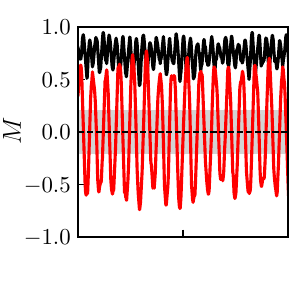}};

\fill[white, draw=black, line width=0.05mm] (12.32-0.18,3.26-0.2) rectangle (12.32+0.18,3.26+0.2);
\fill[white, draw=black, line width=0.05mm] (12.38-0.25,1.35-0.2) rectangle (12.38+0.25,1.35+0.2);
\fill[white, draw=black, line width=0.05mm] (12.34-0.21,-0.56-0.2) rectangle (12.34+0.21,-0.56+0.2);

\node[rotate=0] at (12.32,3.27) {\small (l)};
\node[rotate=0] at (12.38,1.36) {\small (m)};
\node[rotate=0] at (12.34,-0.55) {\small (n)};

\node[rotate=0, anchor=east] at (13.78,3.15) {\scriptsize \textbf{WW}};
\node[rotate=0, anchor=east] at (13.78,1.24) {\scriptsize \textbf{WW}};
\node[rotate=0, anchor=east] at (13.78,-0.67) {\scriptsize \textbf{NICA}};

\node[rotate=0] at (15.45,0.61) {\includegraphics[height=2.9cm]{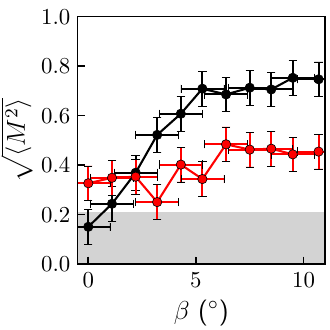}};
\node[rotate=0] at (15.45,2.99) {\includegraphics[height=2.9cm]{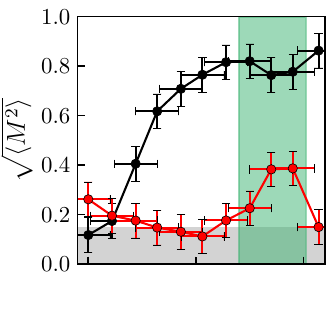}};

\node[rotate=0] at (14.93,4.01) {\small (o)};
\node[rotate=0] at (14.93,1.64) {\small (p)};

\end{tikzpicture}
\vspace*{-0.85cm}
\caption{\small{Collective actuation in the presence of an external field. (a/c/e) Real space dynamics (same color code and scale bars as in Fig. \ref{fig:fig2}-d) and (b/d/f) polarizations in the transverse (red) and longitudinal (black) direction for a triangular lattice pinned at its edges, for increasing field; (bottom) $h = 0$ ($\beta = 0^{\circ}$), CO; (middle) $h = 0.11$ ($\beta = 10.7^{\circ}$), WW; (top) $h = 0.21$ ($\beta = 21.4^{\circ}$), FP; (g/k) Real space dynamics for square lattices pinned at both ends (top and bottom rows are pinned) for $\beta = 8.5^{\circ}$, with $W=4$ (g) and $W=2$ (k); and (h-j) and (l-n) same as (b/d/f) for respectively the stiff ($W = 4$) and soft ($W = 2$) lattices for $h = 0$ ($\beta = 0^{\circ}$) (bottom), $h = 8.7 \times 10^{-4}$ ($\beta = 8.5^{\circ}$) (middle), $h = 1.1 \times 10^{-3}$ ($\beta = 10.7^{\circ}$) (top). (o-p) Root mean squared transverse (red) and longitudinal (black) polarizations as a function of the tilt $\beta$ for $W = 4$ (o), and $W = 2$ (p). The gray areas cover the range of polarization accessible for a system of $N$ randomly oriented unit vectors ($[-1/\sqrt{N}, 1/\sqrt{N}]$ for $M_{\perp/\parallel}$ and $[0, 1/\sqrt{N}]$ for $\langle M_{\perp/\parallel}^2 \rangle^{1/2}$); the green area indicates the region where the reentrance transition to a WW regime occurs.}}
\vspace{-0.5cm}
\label{fig:fig3}
\end{figure*}

Moving to collective dynamics, we start with a triangular lattice of soft springs with $N = 19$ nodes pinned at its edges. This geometry has been explored in the zero-field case and a transition to collective CO was reported~\cite{baconnier2022selective}.
The dynamics condensates on the subspace spanned by the two modes with the degenerate lowest eigenvalue $\omega_{0}^{2}$ of the dynamical matrix, and their harmonics.
The threshold for CO scales as $\Pi\sim\omega_0^2$. 
We set $\Pi = 1.92 \omega_{0}^{2}$, for which CO was observed when $h = 0$, and measure the longitudinal and transverse polarizations $M_{\parallel,\perp}(t) = (1/N) \sum_i \boldsymbol{\hat{n}}_i(t) \cdot \boldsymbol{\hat{e}}_{\parallel,\perp}$ for several field amplitudes.
For small enough fields, the CO regime is preserved; the system is polarized, and this polarization rotates in time, so that the transverse and longitudinal polarizations oscillate in quadrature (Figs. \ref{fig:fig3}e-f, Movie 4). Increasing $h$, the CO regime is replaced by a collective WW regime in which the system polarizes longitudinally and the transverse polarization oscillates, yet with a smaller frequency than in the CO regime (Figs. \ref{fig:fig3}c-d, Movie 5).
The longitudinal polarization is large and never changes sign, while being modulated at twice the frequency of the transverse oscillations.
For even larger fields, the system freezes in the FP regime, with a longitudinal polarization close to one (Figs. \ref{fig:fig3}a-b, Movie 6).
Altogether, we recover the same transition sequence as for the single active unit case.

Changing the geometry allows to consider a gapped system where the longitudinal direction is much stiffer than the perpendicular one.
We consider two square lattices of stiff springs, composed of $L = 12$ (resp. $W = 2$ or $4$) active units along the long (resp. short) direction, pinned at both ends in the long direction~ (Figs. \ref{fig:fig3}g-k).
This geometry has been investigated in the zero-field case~ \cite{baconnier2024noise}.
For the stiffer network, $W = 4$, the system is Frozen-Disordered (FD) (Fig. \ref{fig:fig3}j): both $M_{\parallel}$ and $M_{\perp}$ are small and fluctuate.
In contrast, for the softer network, $W = 2$, the transverse polarization $M_{\perp}$ oscillates while the longitudinal one $M_{\parallel}$ remains small and noisy (Fig. \ref{fig:fig3}n):
this is a NICA regime.
Imposing a field along the stiff direction, the NICA regime turns into a WW regime (Figs. \ref{fig:fig3}l-m, Movie 7), which is anticipated to give way to the FP regime at even larger fields (not accessible experimentally).
More intriguing is the case of the stiffer network: imposing an increasing field, the FD regime first polarizes in a FP regime as expected; yet, increasing further $h$, a WW regime emerges in an intermediate range of field (Fig. \ref{fig:fig3}i), before stabilizing into the FP regime at larger field (Fig. \ref{fig:fig3}h).
This is illustrated by monitoring the root mean square transverse polarization $\langle M_{\perp}^2 \rangle^{1/2}$ as a function of the tilting angle $\beta$ (Figs. \ref{fig:fig3}o-p).
It implies that the threshold in $\Pi$ above which collective actuation emerges decreases as the system is polarized perpendicularly to the oscillating mode, leading to a reentrant transition to collective actuation.

\begin{figure}[t!]
\centering
\hspace*{-0.45cm}
\begin{tikzpicture}

\node[rotate=0] at (-10.65,3.05) {\includegraphics[width=1.6cm]{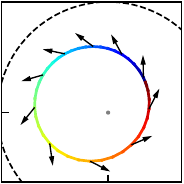}};

\node[rotate=90] at (-11.65,3.05) {\scriptsize $y$};
\node[rotate=0] at (-10.65,2.07) {\scriptsize $x$};

\node[rotate=0] at (-8.15,3.05) {\includegraphics[width=1.6cm]{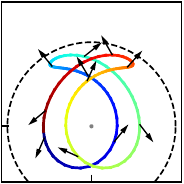}};

\node[rotate=90] at (-9.15,3.05) {\scriptsize $y$};
\node[rotate=0] at (-8.15,2.07) {\scriptsize $x$};

\node[rotate=0] at (-5.7,3.05) {\includegraphics[width=1.6cm]{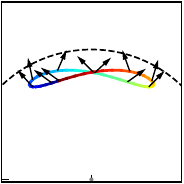}};

\node[rotate=90] at (-6.7,3.05) {\scriptsize $y$};
\node[rotate=0] at (-5.7,2.07) {\scriptsize $x$};

\node[rotate=0, anchor=east] at (-9.83,3.63) {\footnotesize \textbf{CO}};
\node[rotate=0, anchor=east] at (-7.33,3.63) {\footnotesize \textbf{WW$^2$}};
\node[rotate=0, anchor=east] at (-4.88,3.63) {\footnotesize \textbf{WW}};

\draw[force] (-4.58,2.75) -- (-4.58,3.4);
\node[rotate=0] at (-4.37,3.05) {\footnotesize \color{myred} $\boldsymbol{h}$};

\node[rotate=0] at (-11.81,3.66) {\small (a)};
\node[rotate=0] at (-9.34,3.66) {\small (b)};
\node[rotate=0] at (-6.86,3.66) {\small (c)};

\node[rotate=0] at (-6.2,0.0) {\includegraphics[width=4.4cm]{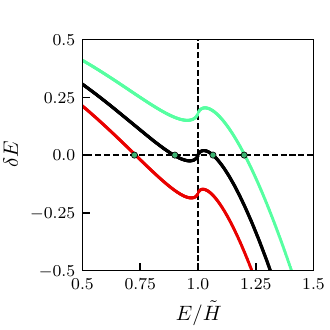}};
\node[rotate=0] at (-6.2,-4.15) {\includegraphics[width=4.4cm]{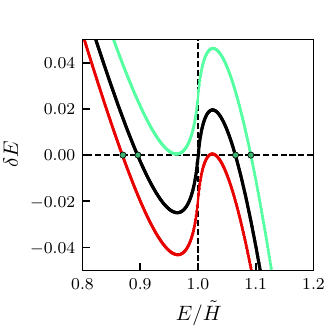}};

\node[rotate=0] at (-4.52,1.38) {\small (f)};
\node[rotate=0] at (-4.52,-2.77) {\small (g)};

\node[rotate=-39] at (-6.75,0.83) {\scriptsize $\tilde{H} = 1/3$};
\node[rotate=-34] at (-6.73,1.25) {\scriptsize \color{colorCO_new} $\tilde{H} = 0.32$};
\node[rotate=-44] at (-6.72,0.05) {\scriptsize \color{colorWW_new} $\tilde{H} = 0.345$};

\node[rotate=-65] at (-6.40,-3.17) {\scriptsize \color{colorCO_new} $\tilde{H} = 0.3314$};
\node[rotate=-65] at (-6.64,-4.72) {\scriptsize \color{colorWW_new} $\tilde{H} = 0.3357$};

\draw[] (-6.3,-0.05) rectangle (-5.25,0.30);
\draw[] (-6.3,-0.05) -- (-7.30,-2.48);
\draw[] (-5.25,-0.05) -- (-4.22,-2.48);

\draw[-{Latex[length=3,width=2]},thin,black] (-5.68,-3.95) to[out=30,in=150] (-5.34,-3.96);
\draw[-{Latex[length=3,width=2]},thin,black] (-5.84,-4.10) to[out=210,in=-30] (-6.47,-4.08);

\node[rotate=0] at (-10.75,0.0) {\includegraphics[width=4.4cm]{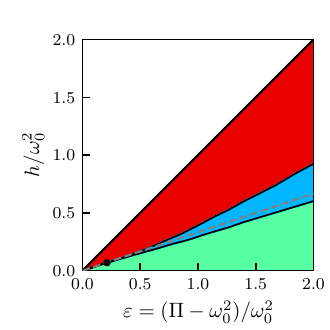}};

\node[rotate=0] at (-11.53,1.38) {\small (d)};

\node[rotate=45, anchor=east] at (-9.3,1.45) {\footnotesize \textbf{FP}};
\node[rotate=45, anchor=east] at (-9.0,1.15) {\footnotesize \textbf{WW}};

\node[rotate=0, anchor=east] at (-8.75,-1.2) {\footnotesize \textbf{CO}};

\node[rotate=33, anchor=east] at (-8.85,0.4) {\footnotesize \textbf{WW$^2$}};
\draw[->] (-9.25,-0.05) -- (-9.15,-0.45);

\node[rotate=0] at (-10.71,-4.15) {\includegraphics[width=4.4cm]{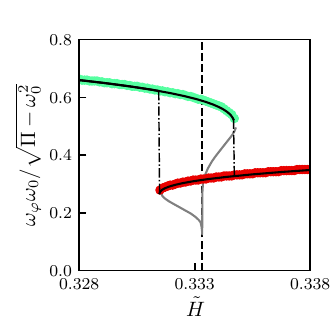}};

\node[rotate=0] at (-11.55,-2.77) {\small (e)};

\node[rotate=0] at (-9.56,-2.76) {\footnotesize $\tilde{H} = 1/3$};

\draw[->, thick, color=black] (-9.63,-3.63) -- (-9.63,-4.13);
\draw[->, thick, color=black] (-10.96,-4.10) -- (-10.96,-3.60);

\node[rotate=5] at (-9.25,-4.5) {\footnotesize \textbf{WW}};
\node[rotate=-9] at (-11.35,-3.32) {\footnotesize \textbf{CO}};

\end{tikzpicture}
\vspace*{-0.88cm}
\caption{\small{Noiseless single active unit, theory and numerics. (a-c) Dynamics of the displacements in the dynamical regimes as named, as obtained from noiseless simulations of Eqs. (\ref{eq:sp}); the trajectory is shown over one period
, and colored with time running from blue to red; the dark arrows are snapshots of the orientation of the active force $\boldsymbol{\hat{n}}$; for $\Pi/\omega_{0}^{2} = 3.95$; (a) $h/\omega_{0}^{2} = 0.62$: CO, (b) $h/\omega_{0}^{2} = 1.15$: WW$^2$, (c) $h/\omega_{0}^{2} = 2.23$: WW; the dashed circles indicate the radius $\Pi/\omega_{0}^{2}$. (d) Numerical phase diagram (white: FP, green: CO, red: WW, blue: WW$^2$); the top solid black (resp. dashed gray) line indicates $\Pi_c = \omega_{0}^{2} + h$ (resp. $\Pi^{\star} = \omega_{0}^{2} + 3h$). The black dot indicates the tip of the WW$^2$ domain of existence.
(e) Rescaled frequency $\omega_{\varphi}$ as a function of $\tilde{H}$, for small $h/\omega_{0}^{2} = 10^{-4}$, colored markers indicate simulations of Eqs. (\ref{eq:sp}) (green: CO, red: WW), and the solid black (resp. gray) lines are the stable (resp. unstable) orbits of the pendulum equations, as obtained from the energy drift $\delta E$.
(f-g) Energy drift $\delta E$ as a function of $E/\tilde{H}$ for three values of $\tilde{H}$, as indicated. The green marks indicate the stable orbits.}}
\label{fig:fig4}
\end{figure}
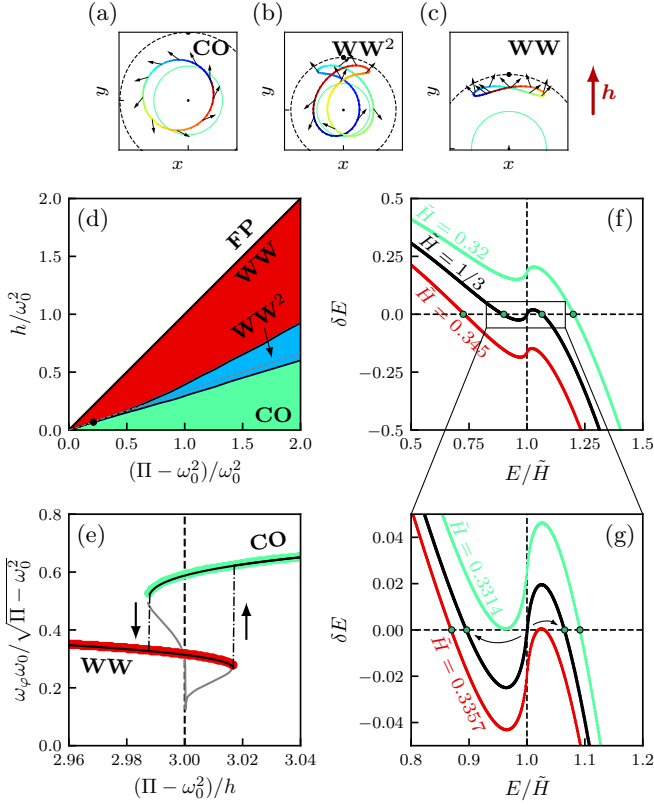

The above results have motivated a systematic study of the effects of a polarizing field on the collective actuation of active elastic structures, which we describe extensively in a companion paper \cite{baconnier2025collective}.
We here summarize the main findings that are relevant for the understanding of the above experiments: we (i) determine analytically $\Pi_c$ and $\Pi^{\star}$ at the single-particle level, and (ii) uncover the origin of the reentrant transition at the collective level.

Integrating numerically the noiseless version of Eqs. ~(\ref{eq:sp}), we obtain the phase diagram shown in Figs. \ref{fig:fig4}a-d.
We recognize the regimes observed experimentally, with the addition of a higher-order windscreen wiper regime WW$^2$. For $h > 0$, there are only two fixed points, with the orientation $\boldsymbol{\hat{n}}$ pointing along or opposite to the field. The latter is always linearly unstable.
The former destabilizes via a supercritical Hopf bifurcation when $\Pi > \Pi_c = \omega_{0}^{2} + h$, leading to the oscillating WW regime, as observed experimentally. Moreover, the eigenvalues and eigenvectors of the Jacobian computed at this fixed point coalesce when $\Pi = \omega_{0}^{2}$ and $h = 0$, indicating an exceptional point, associated with the rotational symmetry and the presence of a Goldstone mode along $\varphi$.

Expanding the dynamics around the exceptional point, one shows that, at zeroth order in $\varepsilon = (\Pi/\omega_0^2) - 1$, the rescaled angular dynamics is described by the equations of motion of a pendulum ($\dot{\tilde{\varphi}} = \tilde{\gamma}$; $\dot{\tilde{\gamma}} = - \tilde{H} \sin \tilde{\varphi}$) with solutions of constant energy $E = \tilde{\gamma}^2 /2 - \tilde{H} \cos \tilde{\varphi}$, where the tildes denote rescaled variables and $\tilde{H} = h/(\varepsilon \omega_{0}^{2})$. In this picture the small, $E < \tilde{H}$, respectively large, $E > \tilde{H}$, energy solutions correspond to the bounded, resp. unbounded, phase dynamics of the pendulum, namely the WW, resp. the CO regimes reported above.
At this order, however, the orbit remains undetermined.
At the next order in $\varepsilon$, an energy drift $\delta E(E, \tilde{H}) \sim \sqrt{\varepsilon}$ drives the system toward different stationary ($\delta E = 0$) and stable ($\partial \delta E/\partial E < 0$) orbits depending on the value of $\tilde{H}$. We find that the CO and WW regimes are selected for $\tilde H\lesssim 1/3$ and $\tilde{H} \gtrsim 1/3$, respectively (Figs. \ref{fig:fig4}f-g).
Moreover, the two regimes coexist within a small range of $\tilde{H}$ close to $\tilde H=1/3$, in agreement with simulations for small $\varepsilon$ (Fig. \ref{fig:fig4}e). Although strictly valid only close to the exceptional point and without noise, this mapping provides a qualitative understanding of the transition between CO and WW regimes, with a threshold $\Pi^{\star}$ in quantitative agreement with the experiments (Fig. \ref{fig:fig2}b).

\begin{figure}[b!]
\centering
\hspace*{-0.2cm}
\begin{tikzpicture}

\node[rotate=0] at (-2.04,3.05) {\includegraphics[width=1.58cm]{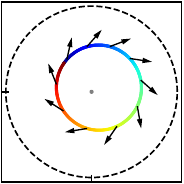}};

\node[rotate=90] at (-3.03,3.05) {\scriptsize $y$};
\node[rotate=0] at (-2.03,2.07) {\scriptsize $x$};

\node[rotate=0] at (-3.19,3.66) {\small (a)};

\node[rotate=0] at (0.44,3.05) {\includegraphics[width=1.58cm]{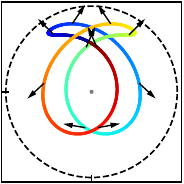}};

\node[rotate=90] at (-0.55,3.05) {\scriptsize $y$};
\node[rotate=0] at (0.45,2.07) {\scriptsize $x$};

\node[rotate=0] at (-0.71,3.66) {\small (b)};

\node[rotate=0] at (2.92,3.05) {\includegraphics[width=1.58cm]{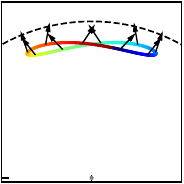}};

\node[rotate=90] at (1.93,3.05) {\scriptsize $y$};
\node[rotate=0] at (2.93,2.07) {\scriptsize $x$};

\node[rotate=0] at (1.76,3.66) {\small (c)};

\draw[force] (4.07,2.75) -- (4.07,3.4);
\node[rotate=0] at (4.28,3.05) {\footnotesize \color{myred} $\boldsymbol{h}$};

\node[rotate=0, anchor=east] at (-1.22,2.46) {\footnotesize \textbf{CO}};
\node[rotate=0, anchor=east] at (1.27,2.48) {\footnotesize \textbf{WW$^2$}};
\node[rotate=0, anchor=east] at (3.72,2.46) {\footnotesize \textbf{WW}};

\node[rotate=0] at (-1.43,0.12) {\includegraphics[width=3.3cm]{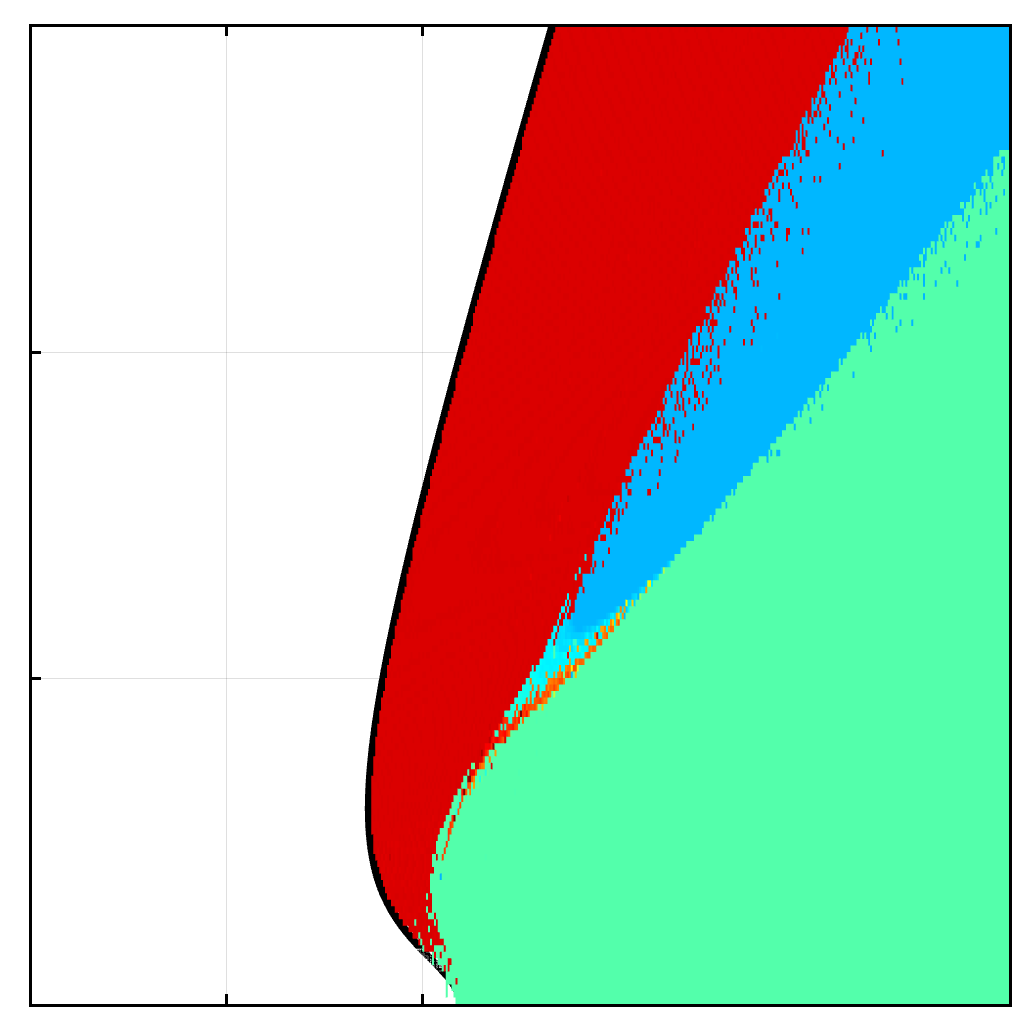}};

\node[rotate=90] at (-3.7,0.12) {\footnotesize $h/\omega_{0}^{2}$};
\node[] at (-1.43,-1.95) {\footnotesize $\Pi/\omega_{0}^{2}$};

\node[] at (-2.97,-1.61) {\scriptsize $0$};
\node[] at (-2.35,-1.61) {\scriptsize $1$};
\node[] at (-1.73,-1.61) {\scriptsize $2$};
\node[] at (-1.11,-1.61) {\scriptsize $3$};
\node[] at (-0.49,-1.61) {\scriptsize $4$};

\node[] at (-3.24,-1.42) {\scriptsize $0.0$};
\node[] at (-3.24,-0.39) {\scriptsize $0.5$};
\node[] at (-3.24,0.64) {\scriptsize $1.0$};

\node[rotate=0] at (2.9,0.12) {\includegraphics[width=3.3cm]{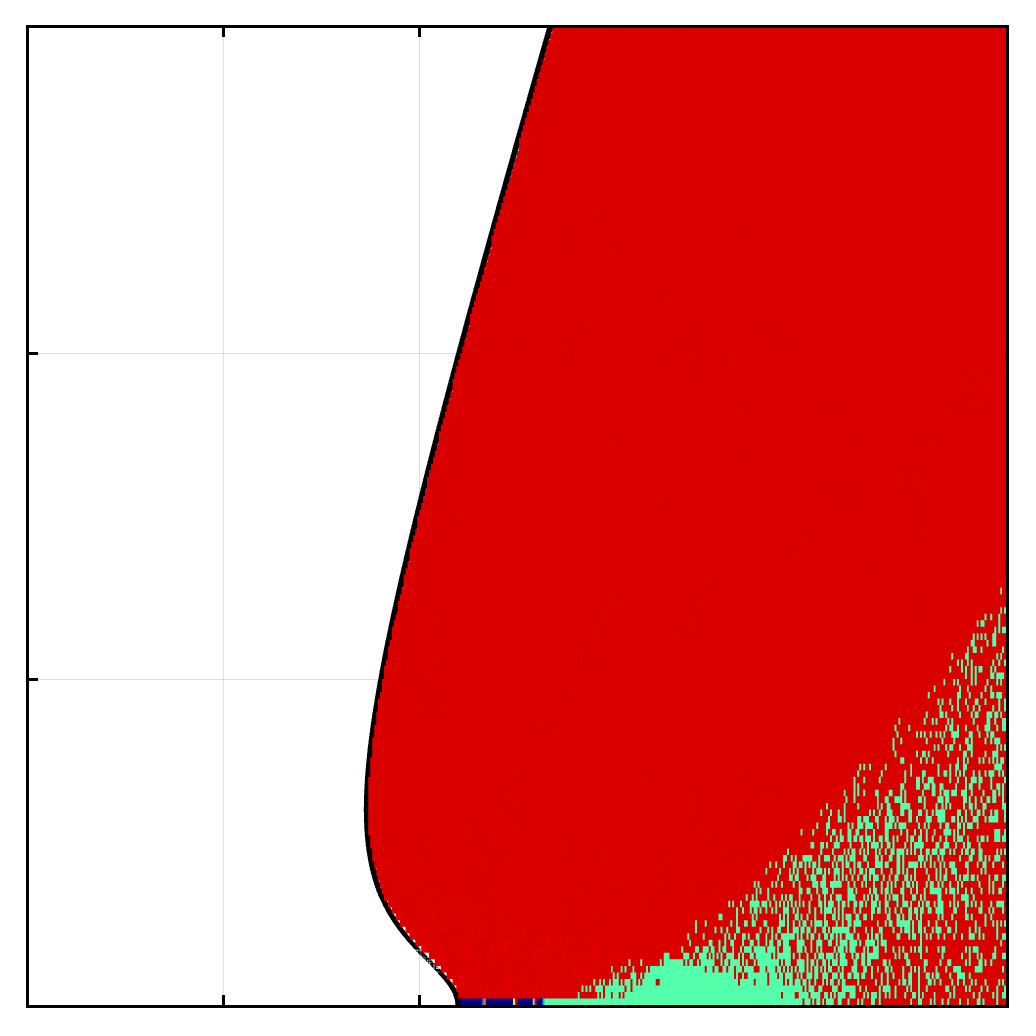}};

\node[rotate=90] at (0.63,0.12) {\footnotesize $h/\omega_{0}^{2}$};
\node[] at (2.9,-1.95) {\footnotesize $\Pi/\omega_{0}^{2}$};

\node[] at (1.36,-1.61) {\scriptsize $0$};
\node[] at (1.98,-1.61) {\scriptsize $1$};
\node[] at (2.6,-1.61) {\scriptsize $2$};
\node[] at (3.22,-1.61) {\scriptsize $3$};
\node[] at (3.84,-1.61) {\scriptsize $4$};

\node[] at (1.09,-1.42) {\scriptsize $0.0$};
\node[] at (1.09,-0.39) {\scriptsize $0.5$};
\node[] at (1.09,0.64) {\scriptsize $1.0$};

\node[rotate=75, anchor=east] at (-1.54,1.67) {\footnotesize \textbf{FP}};
\node[rotate=75, anchor=east] at (-1.13,1.64) {\footnotesize \textbf{WW}};
\node[rotate=0, anchor=east] at (0.17,-1.23) {\footnotesize \textbf{CO}};

\node[rotate=0] at (-2.27,0.65) {\footnotesize \textbf{WW}$^2$};
\draw[] (-2.57,0.45) -- (-2.07,0.45);
\draw[->] (-2.32,0.45) -- (-0.97,0.05);

\node[rotate=75, anchor=east] at (2.79,1.67) {\footnotesize \textbf{FP}};
\node[rotate=75, anchor=east] at (3.20,1.64) {\footnotesize \textbf{WW}};

\node[rotate=0] at (-2.68,1.4) {\small (d)};
\node[rotate=0] at (1.65,1.4) {\small (e)};

\node[] at (3.04,-0.57) {\footnotesize \textbf{NICA}};
\draw[] (2.74,-0.75) -- (3.34,-0.75);
\draw[->] (3.04,-0.75) -- (2.87,-1.37);

\draw[] (2.85,-1.43) ellipse (0.2 and 0.09);

\node[rotate=0] at (3.8,-0.08) {\footnotesize \textbf{CO}};
\draw[] (3.6,-0.26) -- (4.0,-0.26);
\draw[->] (3.8,-0.26) -- (3.4,-1.40);

\end{tikzpicture}
\vspace*{-0.9cm}
\caption{\small{Coarse-grained description, theory and numerics. (a-c) Dynamics of the displacements in the dynamical regimes as named, in the degenerate case, obtained by simulating Eqs. (\ref{eq:cg}); same conventions as Figs. \ref{fig:fig4}a-c. (d-e) Phase diagrams in the degenerate case (d), $\omega_{\parallel}^{2} = \omega_{\perp}^{2} = \omega_{0}^{2}$, and the non-degenerate case (e), $\omega_{\parallel}^{2} = 2\omega_{\perp}^{2} = 2\omega_{0}^{2}$; same color code as Fig. \ref{fig:fig4}d, dark blue markers at $h = 0$ indicate the NICA regime; the black lines represent $\Pi_c(h,D)$ as given by Eq. (\ref{eq:threshold}); areas where two colors appear intermingled indicate zones of coexistence between regimes. In all panels, $D=0.1$ and $\omega_{0}^{2} = 1$ are fixed.}}
\label{fig:fig5}
\end{figure}

Finally, we discuss the generalization of the coarse-grained description of the large-scale displacement $\boldsymbol{U}(\boldsymbol{r}, t)$ and polarization $\boldsymbol{m}(\boldsymbol{r}, t)$ fields introduced in \cite{baconnier2022selective} to the presence of a polarizing field \cite{baconnier2025collective}. The mean-field-like equations are written for a dynamics projected on two spatially homogeneous modes, the longitudinal and transverse modes, of stiffnesses $\omega_{\parallel}^{2}$ and $\omega_{\perp}^{2}$ respectively:
\begin{subequations} \label{eq:cg}
\begin{align}
\dot{\boldsymbol{U}} &= \Pi \boldsymbol{m} - \mathbb{M} \boldsymbol{U}, \label{eq1:cg} \\
\dot{\boldsymbol{m}} &= \left(\!\boldsymbol{m}\!\times\!\left(\!\dot{\boldsymbol{U}} \!+\! \boldsymbol{h} \!\right)\!\right)\!\times\!\boldsymbol{m} + \frac{1 \!-\! \boldsymbol{m}^2}{2} \left(\!\dot{\boldsymbol{U}} \!+\! \boldsymbol{h} \!\right) \!-\! D \boldsymbol{m}, \label{eq2:cg}
\end{align}
\end{subequations}
where the matrix $\mathbb{M}$ is diagonal with entries $\omega_{\parallel}^{2}$ and $\omega_{\perp}^{2}$, and $D$ is the rotational diffusion coefficient of the orientations.
The dynamics has a single fixed point corresponding to a frozen system, polarized in the direction of the field.
In the degenerate case, $\omega_{\parallel}^{2} = \omega_{\perp}^{2}$, or when the field is imposed along the stiff direction, i.e. $\omega_{\perp}^{2} < \omega_{\parallel}^{2}$, the linear stability threshold of the fixed point reads
\begin{equation} \label{eq:threshold}
 \Pi_c (h, D) = \frac{ h^2 \left( \omega_{\perp}^{2} + \sqrt{ D^2 + h^2} \right) }{ D^2 + h^2 - D \sqrt{ D^2 + h^2} }.
\end{equation}
For low enough noise, $D < \omega_{\perp}^2$, the non-monotonic dependence of $\Pi_c (h, D)$ with $h$ translates into a reentrant transition to collective actuation (black lines in Figs. \ref{fig:fig5}d-e).
Simulations of Eqs.~(\ref{eq:cg}) also allow to recover the observed periodic orbits.
In the degenerate case (Fig. \ref{fig:fig5}a-d), one easily identifies the CO, WW$^2$, WW, and FP regimes.
The boundaries between these regimes reveals coexistence dynamics, which we further discuss in our companion paper \cite{baconnier2025collective}. For the gapped system, we consider $\omega_{\parallel}^{2} = 2 \omega_{\perp}^2$, and find that the NICA regime is replaced by the WW regime for any $h > 0$ (Fig. \ref{fig:fig5}e).
For large enough $\Pi$ and not too large $h>0$, we observe a wide domain of coexistence between CO and WW. In all cases, for large enough fields, the FP regime is stabilized.

Here, we have shown that applying an external field on an active solid leads to the emergence of new self-oscillating dynamics and substantially affects the transition to collective actuation. Two important challenges lie ahead when trying to apply those ideas to living systems. A key hurdle is to identify and tune the coupling between the field and the living units, e.g., through chemotaxis or galvanotaxis in the context of cell migration~\cite{sengupta2021principles}. Also the strong nonlinear nature of the dynamics promoted by the coupling makes it challenging to tackle analytically and truly control.

\begin{acknowledgments}

P. B. was supported by a Ph.D. grant from ED564 ``Physique en Ile de France”.

\end{acknowledgments}

\bibliography{bibli.bib}

\begin{thebibliography}{46}
\expandafter\ifx\csname natexlab\endcsname\relax\def\natexlab#1{#1}\fi
\expandafter\ifx\csname bibnamefont\endcsname\relax
  \def\bibnamefont#1{#1}\fi
\expandafter\ifx\csname bibfnamefont\endcsname\relax
  \def\bibfnamefont#1{#1}\fi
\expandafter\ifx\csname citenamefont\endcsname\relax
  \def\citenamefont#1{#1}\fi
\expandafter\ifx\csname url\endcsname\relax
  \def\url#1{\texttt{#1}}\fi
\expandafter\ifx\csname urlprefix\endcsname\relax\def\urlprefix{URL }\fi
\providecommand{\bibinfo}[2]{#2}
\providecommand{\eprint}[2][]{\url{#2}}

\bibitem[{\citenamefont{Koenderink et~al.}(2009)\citenamefont{Koenderink,
  Dogic, Nakamura, Bendix, MacKintosh, Hartwig, Stossel, and
  Weitz}}]{koenderink2009active}
\bibinfo{author}{\bibfnamefont{G.~H.} \bibnamefont{Koenderink}},
  \bibinfo{author}{\bibfnamefont{Z.}~\bibnamefont{Dogic}},
  \bibinfo{author}{\bibfnamefont{F.}~\bibnamefont{Nakamura}},
  \bibinfo{author}{\bibfnamefont{P.~M.} \bibnamefont{Bendix}},
  \bibinfo{author}{\bibfnamefont{F.~C.} \bibnamefont{MacKintosh}},
  \bibinfo{author}{\bibfnamefont{J.~H.} \bibnamefont{Hartwig}},
  \bibinfo{author}{\bibfnamefont{T.~P.} \bibnamefont{Stossel}},
  \bibnamefont{and} \bibinfo{author}{\bibfnamefont{D.~A.} \bibnamefont{Weitz}},
  \bibinfo{journal}{Proc. Natl Acad. Sci.} \textbf{\bibinfo{volume}{106}},
  \bibinfo{pages}{15192} (\bibinfo{year}{2009}).

\bibitem[{\citenamefont{Serra-Picamal et~al.}(2012)\citenamefont{Serra-Picamal,
  Conte, Vincent, Anon, Tambe, Bazellieres, Butler, Fredberg, and
  Trepat}}]{serra2012mechanical}
\bibinfo{author}{\bibfnamefont{X.}~\bibnamefont{Serra-Picamal}},
  \bibinfo{author}{\bibfnamefont{V.}~\bibnamefont{Conte}},
  \bibinfo{author}{\bibfnamefont{R.}~\bibnamefont{Vincent}},
  \bibinfo{author}{\bibfnamefont{E.}~\bibnamefont{Anon}},
  \bibinfo{author}{\bibfnamefont{D.~T.} \bibnamefont{Tambe}},
  \bibinfo{author}{\bibfnamefont{E.}~\bibnamefont{Bazellieres}},
  \bibinfo{author}{\bibfnamefont{J.~P.} \bibnamefont{Butler}},
  \bibinfo{author}{\bibfnamefont{J.~J.} \bibnamefont{Fredberg}},
  \bibnamefont{and} \bibinfo{author}{\bibfnamefont{X.}~\bibnamefont{Trepat}},
  \bibinfo{journal}{Nature Physics} \textbf{\bibinfo{volume}{8}},
  \bibinfo{pages}{628} (\bibinfo{year}{2012}).

\bibitem[{\citenamefont{Deforet et~al.}(2014)\citenamefont{Deforet, Hakim,
  Yevick, Duclos, and Silberzan}}]{deforet2014emergence}
\bibinfo{author}{\bibfnamefont{M.}~\bibnamefont{Deforet}},
  \bibinfo{author}{\bibfnamefont{V.}~\bibnamefont{Hakim}},
  \bibinfo{author}{\bibfnamefont{H.~G.} \bibnamefont{Yevick}},
  \bibinfo{author}{\bibfnamefont{G.}~\bibnamefont{Duclos}}, \bibnamefont{and}
  \bibinfo{author}{\bibfnamefont{P.}~\bibnamefont{Silberzan}},
  \bibinfo{journal}{Nature communications} \textbf{\bibinfo{volume}{5}},
  \bibinfo{pages}{3747} (\bibinfo{year}{2014}).

\bibitem[{\citenamefont{Prost et~al.}(2015)\citenamefont{Prost, J{\"u}licher,
  and Joanny}}]{prost2015active}
\bibinfo{author}{\bibfnamefont{J.}~\bibnamefont{Prost}},
  \bibinfo{author}{\bibfnamefont{F.}~\bibnamefont{J{\"u}licher}},
  \bibnamefont{and} \bibinfo{author}{\bibfnamefont{J.-F.}
  \bibnamefont{Joanny}}, \bibinfo{journal}{Nat. Phys.}
  \textbf{\bibinfo{volume}{11}}, \bibinfo{pages}{111} (\bibinfo{year}{2015}).

\bibitem[{\citenamefont{Bull et~al.}(2021)\citenamefont{Bull, Prakash, and
  Prakash}}]{bull2021ciliary}
\bibinfo{author}{\bibfnamefont{M.~S.} \bibnamefont{Bull}},
  \bibinfo{author}{\bibfnamefont{V.~N.} \bibnamefont{Prakash}},
  \bibnamefont{and} \bibinfo{author}{\bibfnamefont{M.}~\bibnamefont{Prakash}},
  \bibinfo{journal}{arXiv preprint arXiv:2107.02934}  (\bibinfo{year}{2021}).

\bibitem[{\citenamefont{Tan et~al.}(2022)\citenamefont{Tan, Mietke, Li, Chen,
  Higinbotham, Foster, Gokhale, Dunkel, and Nikta}}]{tan2022odd}
\bibinfo{author}{\bibfnamefont{T.~H.} \bibnamefont{Tan}},
  \bibinfo{author}{\bibfnamefont{A.}~\bibnamefont{Mietke}},
  \bibinfo{author}{\bibfnamefont{J.}~\bibnamefont{Li}},
  \bibinfo{author}{\bibfnamefont{Y.}~\bibnamefont{Chen}},
  \bibinfo{author}{\bibfnamefont{H.}~\bibnamefont{Higinbotham}},
  \bibinfo{author}{\bibfnamefont{P.~J.} \bibnamefont{Foster}},
  \bibinfo{author}{\bibfnamefont{S.}~\bibnamefont{Gokhale}},
  \bibinfo{author}{\bibfnamefont{J.}~\bibnamefont{Dunkel}}, \bibnamefont{and}
  \bibinfo{author}{\bibfnamefont{F.}~\bibnamefont{Nikta}},
  \bibinfo{journal}{Nature} \textbf{\bibinfo{volume}{607}},
  \bibinfo{pages}{287} (\bibinfo{year}{2022}).

\bibitem[{\citenamefont{Ferrante et~al.}(2013)\citenamefont{Ferrante, Turgut,
  Dorigo, and Huepe}}]{ferrante2013elasticity}
\bibinfo{author}{\bibfnamefont{E.}~\bibnamefont{Ferrante}},
  \bibinfo{author}{\bibfnamefont{A.~E.} \bibnamefont{Turgut}},
  \bibinfo{author}{\bibfnamefont{M.}~\bibnamefont{Dorigo}}, \bibnamefont{and}
  \bibinfo{author}{\bibfnamefont{C.}~\bibnamefont{Huepe}},
  \bibinfo{journal}{Phys. Rev. Lett.} \textbf{\bibinfo{volume}{111}},
  \bibinfo{pages}{268302} (\bibinfo{year}{2013}).

\bibitem[{\citenamefont{Briand and Dauchot}(2016)}]{briand2016crystallization}
\bibinfo{author}{\bibfnamefont{G.}~\bibnamefont{Briand}} \bibnamefont{and}
  \bibinfo{author}{\bibfnamefont{O.}~\bibnamefont{Dauchot}},
  \bibinfo{journal}{Phys. Rev. Lett.} \textbf{\bibinfo{volume}{117}},
  \bibinfo{pages}{098004} (\bibinfo{year}{2016}).

\bibitem[{\citenamefont{Briand et~al.}(2018)\citenamefont{Briand, Schindler,
  and Dauchot}}]{briand2018spontaneously}
\bibinfo{author}{\bibfnamefont{G.}~\bibnamefont{Briand}},
  \bibinfo{author}{\bibfnamefont{M.}~\bibnamefont{Schindler}},
  \bibnamefont{and} \bibinfo{author}{\bibfnamefont{O.}~\bibnamefont{Dauchot}},
  \bibinfo{journal}{Phys. Rev. Lett.} \textbf{\bibinfo{volume}{120}},
  \bibinfo{pages}{208001} (\bibinfo{year}{2018}).

\bibitem[{\citenamefont{Li et~al.}(2019)\citenamefont{Li, Batra, Brown, Chang,
  Ranganathan, Hoberman, Rus, and Lipson}}]{li2019particle}
\bibinfo{author}{\bibfnamefont{S.}~\bibnamefont{Li}},
  \bibinfo{author}{\bibfnamefont{R.}~\bibnamefont{Batra}},
  \bibinfo{author}{\bibfnamefont{D.}~\bibnamefont{Brown}},
  \bibinfo{author}{\bibfnamefont{H.-D.} \bibnamefont{Chang}},
  \bibinfo{author}{\bibfnamefont{N.}~\bibnamefont{Ranganathan}},
  \bibinfo{author}{\bibfnamefont{C.}~\bibnamefont{Hoberman}},
  \bibinfo{author}{\bibfnamefont{D.}~\bibnamefont{Rus}}, \bibnamefont{and}
  \bibinfo{author}{\bibfnamefont{H.}~\bibnamefont{Lipson}},
  \bibinfo{journal}{Nature} \textbf{\bibinfo{volume}{567}},
  \bibinfo{pages}{361} (\bibinfo{year}{2019}).

\bibitem[{\citenamefont{Zheng et~al.}(2023)\citenamefont{Zheng, Brandenbourger,
  Robinet, Schall, Lerner, and Coulais}}]{zheng2023self}
\bibinfo{author}{\bibfnamefont{E.}~\bibnamefont{Zheng}},
  \bibinfo{author}{\bibfnamefont{M.}~\bibnamefont{Brandenbourger}},
  \bibinfo{author}{\bibfnamefont{L.}~\bibnamefont{Robinet}},
  \bibinfo{author}{\bibfnamefont{P.}~\bibnamefont{Schall}},
  \bibinfo{author}{\bibfnamefont{E.}~\bibnamefont{Lerner}}, \bibnamefont{and}
  \bibinfo{author}{\bibfnamefont{C.}~\bibnamefont{Coulais}},
  \bibinfo{journal}{Phys. Rev. Lett.} \textbf{\bibinfo{volume}{130}},
  \bibinfo{pages}{178202} (\bibinfo{year}{2023}).

\bibitem[{\citenamefont{Xi et~al.}(2024)\citenamefont{Xi, Marzin, Huang, Jones,
  and Brun}}]{xi2024emergent}
\bibinfo{author}{\bibfnamefont{Y.}~\bibnamefont{Xi}},
  \bibinfo{author}{\bibfnamefont{T.}~\bibnamefont{Marzin}},
  \bibinfo{author}{\bibfnamefont{R.~B.} \bibnamefont{Huang}},
  \bibinfo{author}{\bibfnamefont{T.~J.} \bibnamefont{Jones}}, \bibnamefont{and}
  \bibinfo{author}{\bibfnamefont{P.-T.} \bibnamefont{Brun}},
  \bibinfo{journal}{Proceedings of the National Academy of Sciences}
  \textbf{\bibinfo{volume}{121}}, \bibinfo{pages}{e2410654121}
  (\bibinfo{year}{2024}).

\bibitem[{\citenamefont{Veenstra et~al.}(2024)\citenamefont{Veenstra, Gamayun,
  Guo, Sarvi, Meinersen, and Coulais}}]{veenstra2024non}
\bibinfo{author}{\bibfnamefont{J.}~\bibnamefont{Veenstra}},
  \bibinfo{author}{\bibfnamefont{O.}~\bibnamefont{Gamayun}},
  \bibinfo{author}{\bibfnamefont{X.}~\bibnamefont{Guo}},
  \bibinfo{author}{\bibfnamefont{A.}~\bibnamefont{Sarvi}},
  \bibinfo{author}{\bibfnamefont{C.~V.} \bibnamefont{Meinersen}},
  \bibnamefont{and} \bibinfo{author}{\bibfnamefont{C.}~\bibnamefont{Coulais}},
  \bibinfo{journal}{Nature} \textbf{\bibinfo{volume}{627}},
  \bibinfo{pages}{528} (\bibinfo{year}{2024}).

\bibitem[{\citenamefont{Veenstra et~al.}(2025)\citenamefont{Veenstra,
  Scheibner, Brandenbourger, Binysh, Souslov, Vitelli, and
  Coulais}}]{veenstra2025adaptive}
\bibinfo{author}{\bibfnamefont{J.}~\bibnamefont{Veenstra}},
  \bibinfo{author}{\bibfnamefont{C.}~\bibnamefont{Scheibner}},
  \bibinfo{author}{\bibfnamefont{M.}~\bibnamefont{Brandenbourger}},
  \bibinfo{author}{\bibfnamefont{J.}~\bibnamefont{Binysh}},
  \bibinfo{author}{\bibfnamefont{A.}~\bibnamefont{Souslov}},
  \bibinfo{author}{\bibfnamefont{V.}~\bibnamefont{Vitelli}}, \bibnamefont{and}
  \bibinfo{author}{\bibfnamefont{C.}~\bibnamefont{Coulais}},
  \bibinfo{journal}{Nature} pp. \bibinfo{pages}{1--7} (\bibinfo{year}{2025}).

\bibitem[{\citenamefont{Henkes et~al.}(2011)\citenamefont{Henkes, Fily, and
  Marchetti}}]{henkes2011active}
\bibinfo{author}{\bibfnamefont{S.}~\bibnamefont{Henkes}},
  \bibinfo{author}{\bibfnamefont{Y.}~\bibnamefont{Fily}}, \bibnamefont{and}
  \bibinfo{author}{\bibfnamefont{M.~C.} \bibnamefont{Marchetti}},
  \bibinfo{journal}{Phys. Rev. E} \textbf{\bibinfo{volume}{84}},
  \bibinfo{pages}{040301} (\bibinfo{year}{2011}).

\bibitem[{\citenamefont{Menzel and L\"owen}(2013)}]{menzel2013traveling}
\bibinfo{author}{\bibfnamefont{A.~M.} \bibnamefont{Menzel}} \bibnamefont{and}
  \bibinfo{author}{\bibfnamefont{H.}~\bibnamefont{L\"owen}},
  \bibinfo{journal}{Phys. Rev. Lett.} \textbf{\bibinfo{volume}{110}},
  \bibinfo{pages}{055702} (\bibinfo{year}{2013}).

\bibitem[{\citenamefont{Berthier and Kurchan}(2013)}]{berthier2013non}
\bibinfo{author}{\bibfnamefont{L.}~\bibnamefont{Berthier}} \bibnamefont{and}
  \bibinfo{author}{\bibfnamefont{J.}~\bibnamefont{Kurchan}},
  \bibinfo{journal}{Nat. Phys.} \textbf{\bibinfo{volume}{9}},
  \bibinfo{pages}{310} (\bibinfo{year}{2013}).

\bibitem[{\citenamefont{Bi et~al.}(2016)\citenamefont{Bi, Yang, Marchetti, and
  Manning}}]{bi2016motility}
\bibinfo{author}{\bibfnamefont{D.}~\bibnamefont{Bi}},
  \bibinfo{author}{\bibfnamefont{X.}~\bibnamefont{Yang}},
  \bibinfo{author}{\bibfnamefont{M.~C.} \bibnamefont{Marchetti}},
  \bibnamefont{and} \bibinfo{author}{\bibfnamefont{M.~L.}
  \bibnamefont{Manning}}, \bibinfo{journal}{Physical Review X}
  \textbf{\bibinfo{volume}{6}}, \bibinfo{pages}{021011} (\bibinfo{year}{2016}).

\bibitem[{\citenamefont{Woodhouse et~al.}(2018)\citenamefont{Woodhouse,
  Ronellenfitsch, and Dunkel}}]{woodhouse2018autonomous}
\bibinfo{author}{\bibfnamefont{F.~G.} \bibnamefont{Woodhouse}},
  \bibinfo{author}{\bibfnamefont{H.}~\bibnamefont{Ronellenfitsch}},
  \bibnamefont{and} \bibinfo{author}{\bibfnamefont{J.}~\bibnamefont{Dunkel}},
  \bibinfo{journal}{Phys. Rev. Lett.} \textbf{\bibinfo{volume}{121}},
  \bibinfo{pages}{178001} (\bibinfo{year}{2018}).

\bibitem[{\citenamefont{Giavazzi et~al.}(2018)\citenamefont{Giavazzi, Paoluzzi,
  Macchi, Bi, Scita, Manning, Cerbino, and Marchetti}}]{giavazzi2018flocking}
\bibinfo{author}{\bibfnamefont{F.}~\bibnamefont{Giavazzi}},
  \bibinfo{author}{\bibfnamefont{M.}~\bibnamefont{Paoluzzi}},
  \bibinfo{author}{\bibfnamefont{M.}~\bibnamefont{Macchi}},
  \bibinfo{author}{\bibfnamefont{D.}~\bibnamefont{Bi}},
  \bibinfo{author}{\bibfnamefont{G.}~\bibnamefont{Scita}},
  \bibinfo{author}{\bibfnamefont{M.~L.} \bibnamefont{Manning}},
  \bibinfo{author}{\bibfnamefont{R.}~\bibnamefont{Cerbino}}, \bibnamefont{and}
  \bibinfo{author}{\bibfnamefont{M.~C.} \bibnamefont{Marchetti}},
  \bibinfo{journal}{Soft Matter} \textbf{\bibinfo{volume}{14}},
  \bibinfo{pages}{3471} (\bibinfo{year}{2018}).

\bibitem[{\citenamefont{Janssen}(2019)}]{janssen2019active}
\bibinfo{author}{\bibfnamefont{L.~M.} \bibnamefont{Janssen}},
  \bibinfo{journal}{Journal of Physics: Condensed Matter}
  \textbf{\bibinfo{volume}{31}}, \bibinfo{pages}{503002}
  (\bibinfo{year}{2019}).

\bibitem[{\citenamefont{Maitra and Ramaswamy}(2019)}]{maitra2019oriented}
\bibinfo{author}{\bibfnamefont{A.}~\bibnamefont{Maitra}} \bibnamefont{and}
  \bibinfo{author}{\bibfnamefont{S.}~\bibnamefont{Ramaswamy}},
  \bibinfo{journal}{Phys. Rev. Lett.} \textbf{\bibinfo{volume}{123}},
  \bibinfo{pages}{238001} (\bibinfo{year}{2019}).

\bibitem[{\citenamefont{Klongvessa et~al.}(2019)\citenamefont{Klongvessa,
  Ginot, Ybert, Cottin-Bizonne, and Leocmach}}]{klongvessa2019active}
\bibinfo{author}{\bibfnamefont{N.}~\bibnamefont{Klongvessa}},
  \bibinfo{author}{\bibfnamefont{F.}~\bibnamefont{Ginot}},
  \bibinfo{author}{\bibfnamefont{C.}~\bibnamefont{Ybert}},
  \bibinfo{author}{\bibfnamefont{C.}~\bibnamefont{Cottin-Bizonne}},
  \bibnamefont{and} \bibinfo{author}{\bibfnamefont{M.}~\bibnamefont{Leocmach}},
  \bibinfo{journal}{Phys. Rev. Lett.} \textbf{\bibinfo{volume}{123}},
  \bibinfo{pages}{248004} (\bibinfo{year}{2019}).

\bibitem[{\citenamefont{Ronceray et~al.}(2019)\citenamefont{Ronceray,
  Broedersz, and Lenz}}]{ronceray2019stress}
\bibinfo{author}{\bibfnamefont{P.}~\bibnamefont{Ronceray}},
  \bibinfo{author}{\bibfnamefont{C.~P.} \bibnamefont{Broedersz}},
  \bibnamefont{and} \bibinfo{author}{\bibfnamefont{M.}~\bibnamefont{Lenz}},
  \bibinfo{journal}{Soft Matter} \textbf{\bibinfo{volume}{15}},
  \bibinfo{pages}{331} (\bibinfo{year}{2019}).

\bibitem[{\citenamefont{Scheibner et~al.}(2020)\citenamefont{Scheibner,
  Souslov, Banerjee, Sur{\'o}wka, Irvine, and Vitelli}}]{scheibner2020odd}
\bibinfo{author}{\bibfnamefont{C.}~\bibnamefont{Scheibner}},
  \bibinfo{author}{\bibfnamefont{A.}~\bibnamefont{Souslov}},
  \bibinfo{author}{\bibfnamefont{D.}~\bibnamefont{Banerjee}},
  \bibinfo{author}{\bibfnamefont{P.}~\bibnamefont{Sur{\'o}wka}},
  \bibinfo{author}{\bibfnamefont{W.~T.} \bibnamefont{Irvine}},
  \bibnamefont{and} \bibinfo{author}{\bibfnamefont{V.}~\bibnamefont{Vitelli}},
  \bibinfo{journal}{Nat. Phys.} \textbf{\bibinfo{volume}{16}},
  \bibinfo{pages}{475} (\bibinfo{year}{2020}).

\bibitem[{\citenamefont{Canavello et~al.}(2024)\citenamefont{Canavello,
  Damascena, Cabral, and de~Souza~Silva}}]{canavello2024polar}
\bibinfo{author}{\bibfnamefont{D.}~\bibnamefont{Canavello}},
  \bibinfo{author}{\bibfnamefont{R.~H.} \bibnamefont{Damascena}},
  \bibinfo{author}{\bibfnamefont{L.~R.} \bibnamefont{Cabral}},
  \bibnamefont{and} \bibinfo{author}{\bibfnamefont{C.~C.}
  \bibnamefont{de~Souza~Silva}}, \bibinfo{journal}{Soft Matter}
  \textbf{\bibinfo{volume}{20}}, \bibinfo{pages}{2310} (\bibinfo{year}{2024}).

\bibitem[{\citenamefont{Petrolli et~al.}(2019)\citenamefont{Petrolli, Le~Goff,
  Tadrous, Martens, Allier, Mandula, Herv{\'e}, Henkes, Sknepnek, Boudou
  et~al.}}]{petrolli2019confinement}
\bibinfo{author}{\bibfnamefont{V.}~\bibnamefont{Petrolli}},
  \bibinfo{author}{\bibfnamefont{M.}~\bibnamefont{Le~Goff}},
  \bibinfo{author}{\bibfnamefont{M.}~\bibnamefont{Tadrous}},
  \bibinfo{author}{\bibfnamefont{K.}~\bibnamefont{Martens}},
  \bibinfo{author}{\bibfnamefont{C.}~\bibnamefont{Allier}},
  \bibinfo{author}{\bibfnamefont{O.}~\bibnamefont{Mandula}},
  \bibinfo{author}{\bibfnamefont{L.}~\bibnamefont{Herv{\'e}}},
  \bibinfo{author}{\bibfnamefont{S.}~\bibnamefont{Henkes}},
  \bibinfo{author}{\bibfnamefont{R.}~\bibnamefont{Sknepnek}},
  \bibinfo{author}{\bibfnamefont{T.}~\bibnamefont{Boudou}},
  \bibnamefont{et~al.}, \bibinfo{journal}{Physical Rev. Lett.}
  \textbf{\bibinfo{volume}{122}}, \bibinfo{pages}{168101}
  (\bibinfo{year}{2019}).

\bibitem[{\citenamefont{Peyret et~al.}(2019)\citenamefont{Peyret, Mueller,
  d’Alessandro, Begnaud, Marcq, M{\`e}ge, Yeomans, Doostmohammadi, and
  Ladoux}}]{peyret2019sustained}
\bibinfo{author}{\bibfnamefont{G.}~\bibnamefont{Peyret}},
  \bibinfo{author}{\bibfnamefont{R.}~\bibnamefont{Mueller}},
  \bibinfo{author}{\bibfnamefont{J.}~\bibnamefont{d’Alessandro}},
  \bibinfo{author}{\bibfnamefont{S.}~\bibnamefont{Begnaud}},
  \bibinfo{author}{\bibfnamefont{P.}~\bibnamefont{Marcq}},
  \bibinfo{author}{\bibfnamefont{R.-M.} \bibnamefont{M{\`e}ge}},
  \bibinfo{author}{\bibfnamefont{J.~M.} \bibnamefont{Yeomans}},
  \bibinfo{author}{\bibfnamefont{A.}~\bibnamefont{Doostmohammadi}},
  \bibnamefont{and} \bibinfo{author}{\bibfnamefont{B.}~\bibnamefont{Ladoux}},
  \bibinfo{journal}{Biophys. J.} \textbf{\bibinfo{volume}{117}},
  \bibinfo{pages}{464} (\bibinfo{year}{2019}).

\bibitem[{\citenamefont{Chen et~al.}(2017)\citenamefont{Chen, Liu, Shi,
  Chat{\'e}, and Wu}}]{chen2017weak}
\bibinfo{author}{\bibfnamefont{C.}~\bibnamefont{Chen}},
  \bibinfo{author}{\bibfnamefont{S.}~\bibnamefont{Liu}},
  \bibinfo{author}{\bibfnamefont{X.-q.} \bibnamefont{Shi}},
  \bibinfo{author}{\bibfnamefont{H.}~\bibnamefont{Chat{\'e}}},
  \bibnamefont{and} \bibinfo{author}{\bibfnamefont{Y.}~\bibnamefont{Wu}},
  \bibinfo{journal}{Nature} \textbf{\bibinfo{volume}{542}},
  \bibinfo{pages}{210} (\bibinfo{year}{2017}).

\bibitem[{\citenamefont{Liu et~al.}(2021)\citenamefont{Liu, Shankar, Marchetti,
  and Wu}}]{liu2021viscoelastic}
\bibinfo{author}{\bibfnamefont{S.}~\bibnamefont{Liu}},
  \bibinfo{author}{\bibfnamefont{S.}~\bibnamefont{Shankar}},
  \bibinfo{author}{\bibfnamefont{M.~C.} \bibnamefont{Marchetti}},
  \bibnamefont{and} \bibinfo{author}{\bibfnamefont{Y.}~\bibnamefont{Wu}},
  \bibinfo{journal}{Nature} \textbf{\bibinfo{volume}{590}}, \bibinfo{pages}{80}
  (\bibinfo{year}{2021}).

\bibitem[{\citenamefont{Gu et~al.}(2025)\citenamefont{Gu, Guiselin, Bain,
  Zuriguel, and Bartolo}}]{Gu2025}
\bibinfo{author}{\bibfnamefont{F.}~\bibnamefont{Gu}},
  \bibinfo{author}{\bibfnamefont{B.}~\bibnamefont{Guiselin}},
  \bibinfo{author}{\bibfnamefont{N.}~\bibnamefont{Bain}},
  \bibinfo{author}{\bibfnamefont{I.}~\bibnamefont{Zuriguel}}, \bibnamefont{and}
  \bibinfo{author}{\bibfnamefont{D.}~\bibnamefont{Bartolo}},
  \bibinfo{journal}{Nature} \textbf{\bibinfo{volume}{638}},
  \bibinfo{pages}{112} (\bibinfo{year}{2025}), ISSN \bibinfo{issn}{1476-4687},
  \urlprefix\url{https://doi.org/10.1038/s41586-024-08514-6}.

\bibitem[{\citenamefont{Baconnier et~al.}(2022)\citenamefont{Baconnier, Shohat,
  Hern{\`a}ndez~L{\`o}pez, Coulais, D{\'e}mery, D{\"u}ring, and
  Dauchot}}]{baconnier2022selective}
\bibinfo{author}{\bibfnamefont{P.}~\bibnamefont{Baconnier}},
  \bibinfo{author}{\bibfnamefont{D.}~\bibnamefont{Shohat}},
  \bibinfo{author}{\bibfnamefont{C.}~\bibnamefont{Hern{\`a}ndez~L{\`o}pez}},
  \bibinfo{author}{\bibfnamefont{C.}~\bibnamefont{Coulais}},
  \bibinfo{author}{\bibfnamefont{V.}~\bibnamefont{D{\'e}mery}},
  \bibinfo{author}{\bibfnamefont{G.}~\bibnamefont{D{\"u}ring}},
  \bibnamefont{and} \bibinfo{author}{\bibfnamefont{O.}~\bibnamefont{Dauchot}},
  \bibinfo{journal}{Nature Physics}  (\bibinfo{year}{2022}).

\bibitem[{\citenamefont{Xu et~al.}(2022)\citenamefont{Xu, Huang, Zhang, and
  Wu}}]{xu2022autonomous}
\bibinfo{author}{\bibfnamefont{H.}~\bibnamefont{Xu}},
  \bibinfo{author}{\bibfnamefont{Y.}~\bibnamefont{Huang}},
  \bibinfo{author}{\bibfnamefont{R.}~\bibnamefont{Zhang}}, \bibnamefont{and}
  \bibinfo{author}{\bibfnamefont{Y.}~\bibnamefont{Wu}}, \bibinfo{journal}{arXiv
  preprint arXiv:2208.09664}  (\bibinfo{year}{2022}).

\bibitem[{\citenamefont{Baconnier et~al.}(2023)\citenamefont{Baconnier, Shohat,
  and Dauchot}}]{baconnier2023discontinuous}
\bibinfo{author}{\bibfnamefont{P.}~\bibnamefont{Baconnier}},
  \bibinfo{author}{\bibfnamefont{D.}~\bibnamefont{Shohat}}, \bibnamefont{and}
  \bibinfo{author}{\bibfnamefont{O.}~\bibnamefont{Dauchot}},
  \bibinfo{journal}{Physical Review Letters} \textbf{\bibinfo{volume}{130}},
  \bibinfo{pages}{028201} (\bibinfo{year}{2023}).

\bibitem[{\citenamefont{Baconnier et~al.}(2024)\citenamefont{Baconnier,
  D{\'e}mery, and Dauchot}}]{baconnier2024noise}
\bibinfo{author}{\bibfnamefont{P.}~\bibnamefont{Baconnier}},
  \bibinfo{author}{\bibfnamefont{V.}~\bibnamefont{D{\'e}mery}},
  \bibnamefont{and} \bibinfo{author}{\bibfnamefont{O.}~\bibnamefont{Dauchot}},
  \bibinfo{journal}{Physical Review E} \textbf{\bibinfo{volume}{109}},
  \bibinfo{pages}{024606} (\bibinfo{year}{2024}).

\bibitem[{\citenamefont{Baconnier
  et~al.}(2025{\natexlab{a}})\citenamefont{Baconnier, Dauchot, D{\'e}mery,
  D{\"u}ring, Henkes, Huepe, and Shee}}]{baconnier2025self}
\bibinfo{author}{\bibfnamefont{P.}~\bibnamefont{Baconnier}},
  \bibinfo{author}{\bibfnamefont{O.}~\bibnamefont{Dauchot}},
  \bibinfo{author}{\bibfnamefont{V.}~\bibnamefont{D{\'e}mery}},
  \bibinfo{author}{\bibfnamefont{G.}~\bibnamefont{D{\"u}ring}},
  \bibinfo{author}{\bibfnamefont{S.}~\bibnamefont{Henkes}},
  \bibinfo{author}{\bibfnamefont{C.}~\bibnamefont{Huepe}}, \bibnamefont{and}
  \bibinfo{author}{\bibfnamefont{A.}~\bibnamefont{Shee}},
  \bibinfo{journal}{Reviews of Modern Physics} \textbf{\bibinfo{volume}{97}},
  \bibinfo{pages}{015007} (\bibinfo{year}{2025}{\natexlab{a}}).

\bibitem[{\citenamefont{SenGupta et~al.}(2021)\citenamefont{SenGupta, Parent,
  and Bear}}]{sengupta2021principles}
\bibinfo{author}{\bibfnamefont{S.}~\bibnamefont{SenGupta}},
  \bibinfo{author}{\bibfnamefont{C.~A.} \bibnamefont{Parent}},
  \bibnamefont{and} \bibinfo{author}{\bibfnamefont{J.~E.} \bibnamefont{Bear}},
  \bibinfo{journal}{Nature Reviews Molecular Cell Biology}
  \textbf{\bibinfo{volume}{22}}, \bibinfo{pages}{529} (\bibinfo{year}{2021}).

\bibitem[{\citenamefont{Kennard and Theriot}(2020)}]{kennard2020osmolarity}
\bibinfo{author}{\bibfnamefont{A.~S.} \bibnamefont{Kennard}} \bibnamefont{and}
  \bibinfo{author}{\bibfnamefont{J.~A.} \bibnamefont{Theriot}},
  \bibinfo{journal}{Elife} \textbf{\bibinfo{volume}{9}} (\bibinfo{year}{2020}).

\bibitem[{\citenamefont{Sun et~al.}(2019)\citenamefont{Sun, Reid, Ferreira,
  Luxardi, Ma, Lokken, Zhu, Xu, Sun, Ryzhuk et~al.}}]{sun2019infection}
\bibinfo{author}{\bibfnamefont{Y.}~\bibnamefont{Sun}},
  \bibinfo{author}{\bibfnamefont{B.}~\bibnamefont{Reid}},
  \bibinfo{author}{\bibfnamefont{F.}~\bibnamefont{Ferreira}},
  \bibinfo{author}{\bibfnamefont{G.}~\bibnamefont{Luxardi}},
  \bibinfo{author}{\bibfnamefont{L.}~\bibnamefont{Ma}},
  \bibinfo{author}{\bibfnamefont{K.~L.} \bibnamefont{Lokken}},
  \bibinfo{author}{\bibfnamefont{K.}~\bibnamefont{Zhu}},
  \bibinfo{author}{\bibfnamefont{G.}~\bibnamefont{Xu}},
  \bibinfo{author}{\bibfnamefont{Y.}~\bibnamefont{Sun}},
  \bibinfo{author}{\bibfnamefont{V.}~\bibnamefont{Ryzhuk}},
  \bibnamefont{et~al.}, \bibinfo{journal}{PLoS biology}
  \textbf{\bibinfo{volume}{17}}, \bibinfo{pages}{e3000044}
  (\bibinfo{year}{2019}).

\bibitem[{\citenamefont{Lecuit and Lenne}(2007)}]{lecuit2007cell}
\bibinfo{author}{\bibfnamefont{T.}~\bibnamefont{Lecuit}} \bibnamefont{and}
  \bibinfo{author}{\bibfnamefont{P.-F.} \bibnamefont{Lenne}},
  \bibinfo{journal}{Nature reviews Molecular cell biology}
  \textbf{\bibinfo{volume}{8}}, \bibinfo{pages}{633} (\bibinfo{year}{2007}).

\bibitem[{\citenamefont{Farge}(2011)}]{farge2011mechanotransduction}
\bibinfo{author}{\bibfnamefont{E.}~\bibnamefont{Farge}},
  \bibinfo{journal}{Current topics in developmental biology}
  \textbf{\bibinfo{volume}{95}}, \bibinfo{pages}{243} (\bibinfo{year}{2011}).

\bibitem[{\citenamefont{Miller and Davidson}(2013)}]{miller2013interplay}
\bibinfo{author}{\bibfnamefont{C.~J.} \bibnamefont{Miller}} \bibnamefont{and}
  \bibinfo{author}{\bibfnamefont{L.~A.} \bibnamefont{Davidson}},
  \bibinfo{journal}{Nature Reviews Genetics} \textbf{\bibinfo{volume}{14}},
  \bibinfo{pages}{733} (\bibinfo{year}{2013}).

\bibitem[{\citenamefont{Goodwin and Nelson}(2021)}]{goodwin2021mechanics}
\bibinfo{author}{\bibfnamefont{K.}~\bibnamefont{Goodwin}} \bibnamefont{and}
  \bibinfo{author}{\bibfnamefont{C.~M.} \bibnamefont{Nelson}},
  \bibinfo{journal}{Developmental cell} \textbf{\bibinfo{volume}{56}},
  \bibinfo{pages}{240} (\bibinfo{year}{2021}).

\bibitem[{\citenamefont{Dauchot and D{\'e}mery}(2019)}]{dauchot2019dynamics}
\bibinfo{author}{\bibfnamefont{O.}~\bibnamefont{Dauchot}} \bibnamefont{and}
  \bibinfo{author}{\bibfnamefont{V.}~\bibnamefont{D{\'e}mery}},
  \bibinfo{journal}{Phys. Rev. Lett.} \textbf{\bibinfo{volume}{122}},
  \bibinfo{pages}{068002} (\bibinfo{year}{2019}).

\bibitem[{\citenamefont{Baconnier
  et~al.}(2025{\natexlab{b}})\citenamefont{Baconnier, D{\'e}mery, and
  Dauchot}}]{baconnier2025collective}
\bibinfo{author}{\bibfnamefont{P.}~\bibnamefont{Baconnier}},
  \bibinfo{author}{\bibfnamefont{V.}~\bibnamefont{D{\'e}mery}},
  \bibnamefont{and} \bibinfo{author}{\bibfnamefont{O.}~\bibnamefont{Dauchot}},
  \bibinfo{journal}{arXiv preprint arXiv:2504.08599}
  (\bibinfo{year}{2025}{\natexlab{b}}).

\bibitem[{\citenamefont{Bain and Bartolo}(2019)}]{bain2019dynamic}
\bibinfo{author}{\bibfnamefont{N.}~\bibnamefont{Bain}} \bibnamefont{and}
  \bibinfo{author}{\bibfnamefont{D.}~\bibnamefont{Bartolo}},
  \bibinfo{journal}{Science} \textbf{\bibinfo{volume}{363}},
  \bibinfo{pages}{46} (\bibinfo{year}{2019}).

\end{thebibliography}

\appendix

\section{Methods}

We use commercial HEXBUG nano\copyright Nitro as in \cite{baconnier2022selective, baconnier2023discontinuous, baconnier2024noise}. We embed these bugs in 3D-printed cylindrical structures of $5$ cm internal diameter, $3$ mm thick, and $14$ mm height (as the hexbugs themselves). These 3D-printed annulus have $4$ or $6$ regularly spaced overhangs, with a central hole to hold the edges of the springs. Moreover, we set a thin PP plastic film on the top of the annulus to restrict the vertical motion of the hexbugs body, which we fix using commercial glue and a 3D-printed $1$ mm thick annulus. These elementary components are connected by coil springs. We use two kinds of springs: stiff springs RSC13 ($k \simeq 100$ N/m, $l_0 \simeq 3$ cm, external diameter $5$ mm) manufactured by Ets. Jean CHAPUIS; and soft springs ($k \simeq 1$ N/m, $l_0 \simeq 8$ cm, external diameter $5$ mm) manufactured by Schweizer Federntechnik. We tune the soft springs stiffness by varying their length, the stiffness $k$ of a coil spring being inversely proportional to $l_0$, all other parameters held constants. In particular, for the single-particle experiments of Fig. 2 of the main text, the spring's lengths are $\left\lbrace\right.\!7.4$, $6.6$, $5.8$, $5.0$, $4.4$, $3.6$, $2.8\!\left.\right\rbrace$ cm. These experiments were conducted with soft springs, with a constant extension $\alpha = 1.16$ (imposing $\omega_{0}^{2} \simeq 1.70$). The amplitude of the external polarizing field writes $h = \mu g \sin \beta$, with $\mu = \zeta D$ and $D = D_{\theta} l_e/v_0$, where $D_{\theta}$ (resp. $v_0$) is the diffusion coefficient of the polarities (resp. the cruise velocity of the hexbugs) \cite{baconnier2022selective}, and where $\zeta$ is a constant characterizing the ratio between the reorientation toward the external field and angular noise. By measuring experimentally the polarization $m$ as a function of the tilt angle $\beta$ for a single particle in a very stiff harmonic potential ($\Pi \ll \omega_{0}^{2}$), we have determined $\zeta \simeq 5$. The experiments of Figs. 3a to f of the main text were conducted with soft springs, with a constant extension $\alpha = 1.29$. The experiments of Figs. 3g to p of the main text were conducted with stiff springs, with a constant extension in the longitudinal direction $\alpha = 1.28$. The tilt of the experiments with respect to the horizontal plane is tuned using a hinge mechanism with a discrete number of possible rest angles. The accessible tilts are $\left\lbrace\right.\!0^{\circ}$, $1.1^{\circ}$, $2.2^{\circ}$, $3.2^{\circ}$, $4.3^{\circ}$, $5.3^{\circ}$, $6.4^{\circ}$, $7.5^{\circ}$, $8.5^{\circ}$, $9.5^{\circ}$, $10.7^{\circ}$, $12.8^{\circ}$, $15.0^{\circ}$, $17.1^{\circ}$, $19.3^{\circ}$, $21.4^{\circ}\!\left.\right\rbrace$. The dynamics of the active elastic structures are captured at $40$ frames per second, and the movies are processed with Python as in \cite{baconnier2022selective, baconnier2023discontinuous, baconnier2024noise}. As the camera attached to the ceiling is held vertically, movies acquired during experiments with a finite tilt $\beta$ must be processed to correct for perspective distortions. This is done using planar homography as in \cite{bain2019dynamic}.

\end{document}